\documentclass[journal=jacsat,manuscript=article]{achemso}
\setkeys{acs}{articletitle = true}

\usepackage{graphicx}
\usepackage{dcolumn}
\usepackage{bm}
\usepackage{graphicx}
\usepackage{caption}
\usepackage{subcaption}
\usepackage[percent]{overpic}  
\usepackage{amsmath}
\usepackage{amssymb}
\usepackage{achemso}
\usepackage{amsmath}
\usepackage{mhchem}
\usepackage{color,soul}
\soulregister\cite7
\soulregister\ref7

\usepackage{xr}
\makeatletter
\newcommand*{\addFileDependency}[1]{
  \typeout{(#1)}
  \@addtofilelist{#1}
  \IfFileExists{#1}{}{\typeout{No file #1.}}
}
\makeatother

\newcommand*{\myexternaldocument}[1]{%
    \externaldocument{#1}%
    \addFileDependency{#1.tex}%
    \addFileDependency{#1.aux}%
}
\myexternaldocument{SI}

\usepackage{tabu}
\usepackage{nth}
\usepackage[dvipsnames]{xcolor}
\definecolor{orange}{rgb}{1.0,0.4,0.0}
\usepackage{MnSymbol}
\usepackage{framed}

\newcommand{\textapprox}{{\raise.17ex\hbox{$\scriptstyle\mathtt{\sim}$}}}

\usepackage{accents}

\title{Unraveling UV Stability in Metal Halide Perovskites: From Degradation Mechanisms to Molecular Passivation}

\author{Xin Wen}
 \affiliation{School of Physical Science and Technology, ShanghaiTech University, Shanghai 201210, China. }

\author{Zhiyi Yao}
 \affiliation{School of Physical Science and Technology, ShanghaiTech University, Shanghai 201210, China. }
 
\author{Wenzhuo Li}
 \affiliation{School of Physical Science and Technology, ShanghaiTech University, Shanghai 201210, China. }

\author{Zhijun Ning}
 \affiliation{School of Physical Science and Technology, ShanghaiTech University, Shanghai 201210, China. }
 
 \author{Fan Zheng}
 \affiliation{School of Physical Science and Technology, ShanghaiTech University, Shanghai 201210, China. }
 \email{zhengfan@shanghaitech.edu.cn}

\begin{document}
\maketitle

\begin{abstract}
Understanding the mechanisms of UV-induced degradation is crucial for enhancing the UV stability of perovskite solar cells. The UV-driven structural dynamics of CH$_3$NH$_3$PbI$_3$ (MAPbI$_3$) are investigated using real-time TDDFT simulations, revealing that under the electron and hole excitation, the distortion of the inorganic framework (PbI) is primarily driven by the electron occupation of Pb-\textit{p} and I-\textit{p} antibonding states, whereas in the hole case, it is mainly governed by the direct cooling induced distortion. We also find that UV accelerates the rotation of MA$^+$ molecules. Further, a BDO molecule is introduced as a passivant, which suppresses structural distortions and provides multi-phonon channels to dissipate carrier cooling energy. Experimental results confirm the UV-protective role of BDO, with suppressed PbI$_2$ formation and improved device stability. These results clarify the mechanism of the UV-induced degradation in the MAPbI$_3$ perovskite and further elucidate how passivation molecules enhance UV stability.
\end{abstract}

\newpage

\section{Introduction}
\indent

In recent years, perovskite solar cells (PSCs) have developed rapidly, with single-junction devices reaching a power conversion efficiency (PCE) of 26.95\% and tandem cells achieving 34.85\%, surpassing the highest efficiency of silicon solar cells\cite{NREL,pvkorigin2009,tandem1,tandem2}. PSCs are considered among the most promising devices for next-generation photovoltaic applications\cite{pvkreview1,pvkreview2}. However, the long-term operational stability of PSCs remains a major challenge for commercialization\cite{commercial1,commercial2}. Perovskite structures are susceptible to degradation under moisture\cite{degradationh2opbi2}, oxygen\cite{degradationo2pbi2}, hot temperatures\cite{degradationtemppbi2}, and ultraviolet (UV) irradiation\cite{degradationuvpbi2}. Stability against moisture and oxygen is effectively improved through encapsulation\cite{encapsulation}. Thermal stability can be enhanced by optimizing device architecture\cite{tempdevicestructure}, improving film morphology\cite{tempmorph}, and incorporating either low-dimensional or mixed-cation perovskite compositions\cite{temp2d,2d,tempmix}. In contrast, UV-induced degradation of PSCs is less well investigated. UV light increases interfacial defects and reduces conductivity in PSCs, and studies have shown that UVb light in the range of 280–320 nm is particularly detrimental to device stability\cite{uvdamage1,uvb}. Strategies to improve UV stability include the incorporation of UV-absorbing materials\cite{uvimprovemolecule}, ion doping\cite{uvimproveinterface-ion}, and interfacial engineering\cite{uvimproveinterface1}. For example, Wang et al. enhanced UV stability by introducing 2-hydroxy-4-methoxybenzophenone, which exhibits strong absorption in the UVb region\cite{uvimprovemolecule-main}. Ouafi et al. reported that Br anion doping in CH$_3$NH$_3$PbI$_3$ (MAPbI$_3$) improves UV stability by stabilizing the cubic phase, which is denser and more robust than the tetragonal phase\cite{uvimproveBr}. Zhu et al. achieved improved UV stability by inserting a 2PACz interlayer between the perovskite and NiO$_x$ layers, effectively passivating interfacial voids and vacancies\cite{uvimprove2PACz}. However, the mechanisms of UV-induced degradation are less investigated and not fully understood. For example, Nickel et al. have observed CH$_3$NH$_2$ and H$_2$ molecule even without the ambient O$_2$. They suspect that the UV light may introduce charge localization on CH$_3$NH$_3^+$ (MA$^+$) molecule. The electron occupies the anti-bonding of NH$_3^+$ and leads to the dissociation of MA$^+$\cite{uvmechanism-o2}. Hang et al. report that MAPbI$_3$ first decomposes into PbI$_2$ under UV illumination due to its intrinsic thermal and photoinstability; then UV-induced electron–hole pairs trigger photocatalytic redox reactions that further generate Pb$^0$ and I$_2$ species\cite{uvmechanism-Sno2}. 

However, identifying the causes of UV-instability and providing a microscopic mechanism in PSCs is still challenging for both experimental measurements and theoretical simulations. In general, the UV-induced structural distortions can be mainly attributed to two mechanisms: (i) The high-energy excited electrons (or holes) may occupy anti-bonding (or bonding) states, which soften the chemical bonds and destablize the structure\cite{occ-chemicalbond}. (ii) During the cooling processes of the excited hot electrons and holes, their potential energies are transformed into the ions' kinetic energy, i.e. increasing temperature of the system and distort the structure\cite{cooling-Si}. More importantly, the second mechanism becomes more intricate at the atomic level. The increasing temperature to structural distortion is not a homogeneous process. Instead, the hot carrier cooling can directly induce various phonon mode excitations depending on the strength of electron-phonon couplings, which will distort the structure quite differently from the uniform thermal effect.
Moreover, this process could be within picosecond or even sub-picosecond. Such ultrafast phonon mode excitation and dacay pose a significant difficulty to experimental measurements in UV-illuminated PSCs which demand high spatial and temporal resolutions\cite{hard,local,coolingps}. Meanwhile, the nature of excited states in UV-instability in PSCs brings difficulties to theoretical simulations, particularly in a complex system with hundreds of atoms for several picoseconds. On one hand, the non-equilibrium hot carrier cooling process must be described correctly by satisfying the detailed balance and considering the decoherence effect\cite{detaileddecoherence}. On the other hand, the ionic potential can not be computed by the ground state molecular dynamics solely, instead, the electronic excitations must be included to illustrate how the movement of ions is altered by the excited carrier and the hot carrier cooling process. Some of these simulations are performed in PSCs, for example, Paillard et al. have used constrained density functional theory (constrained-DFT) to enforce some electrons (holes) in the conduction (valence) bands and find that these carriers can reduce the relative energy difference between the \(\delta\) phase and the \(\gamma\) perovskite phase of CsPbI\(_3\)\cite{example-cdft}; Banerjee et al. employ a nonadiabatic molecular dynamics (NAMD) scheme with detailed balance and decoherence effect to describe hot carrier cooling in MAPbI\(_3\), yet the nuclei remain on a ground-state trajectory due to the classical path approximation\cite{example-namd}.

Recently, efficient real-time time-dependent density functional theory (rt-TDDFT) methods have been developed to study the light-excitation\cite{TDDFT-exciton}, hot carrier cooling\cite{TDDFT-cooling}, light-induced phase transition\cite{TDDFT-phase}, and other electron-ion coupled problems\cite{TDDFT-other1,TDDFT-review}. By imposing the detailed balance and conserving the total energy, the hot carrier cooling process in a periodic system and the ionic dynamics on an excited potential for a relatively long time (e.g. several picoseconds) can be described accurately and efficiently. In this work, by utilizing the rt-TDDFT, we have studied the UV-induced hot carrier cooling and the structural distortion in MAPbI$_3$ surface. Instead of using the conventional Ehrenfest rt-TDDFT, a so-called natural orbital branching (TDDFT-NOB) formalism is used to correct the averaged excited-state potential for ions and introduce the  stochastic behavior\cite{nob}. In this method, the Hamiltonian is determined self-consistently to allow for a large time step of TDDFT around 0.1 fs. By using the adiabatic states as the basis to evolve the electronic wavefunctions, it is possible to simulated the electron and ions' dynamics for a system with hundreds of atoms for several picoseconds. To conserve the total energy and satisfy the detailed balance, the electronic potential energy change induced by the hot carrier cooling and the wavefunction collapsing during the NOB event is compensated by reassigning the kinetic energy of ions. Here, the transition degree of freedom (indicating the non-adiabatic coupling direction) is computed on the fly, where the electron-phonon coupling strength is taking into account while adjusting the ions' kinetic energy. By using this method, we have found that the UV-induced hot carriers can distort the MAPbI$_3$ structure significantly. However, these distortions are not a general consequence of thermal effect. Instead, both the excited electron/hole occupations of anti-bonding/bonding state and the direct hot carrier cooling induced structural change are contributing to the UV-instability. More interestingly, although the excited electron and hole are generating similar distortions, their mechanisms are different. Based on the mechanisms found in our calculation, we have proposed to suppress the UV-instability by passivating the MAPbI$_3$ surface with specific molecules. A further experimental verification using the Grazing-Incidence Wide-Angle X-ray Scattering (GIWAXS) is performed to demonstrate the structural instability after UV illumination in the bare MAPbI$_3$. The enhanced stability by the passivation molecule is also experimentally examined and proved its validity.

\section{Results and discussion}
\indent

\begin{figure}[H]
\centering
\includegraphics[width=1\textwidth]{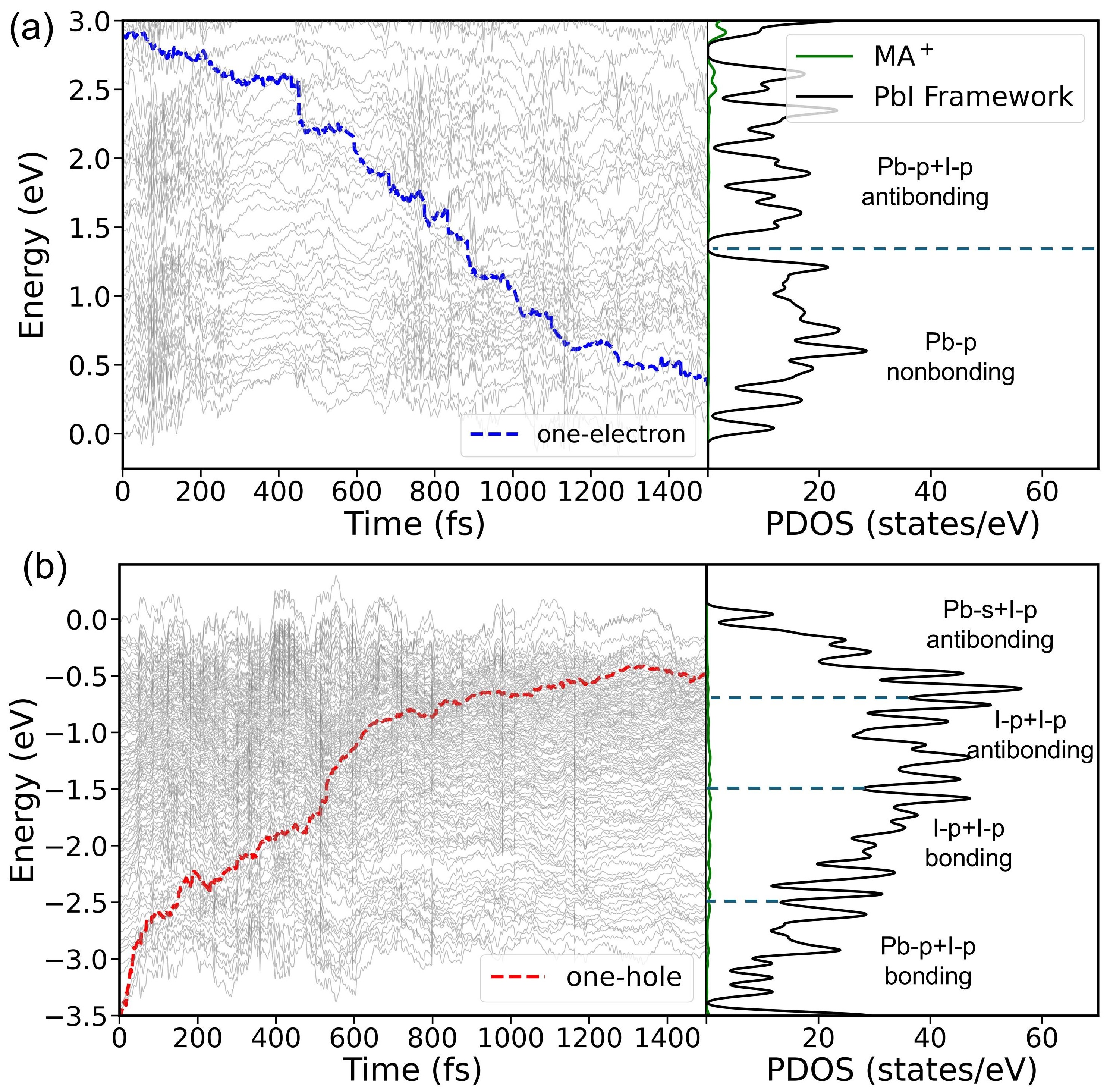}
\captionof{figure}{(a) Electron excitation: Time evolution of eigenenergies in the conduction band (gray lines); the blue line denotes the average energy of the excited electron (the averaged energy of carrier is computed with Equ.S4). (b) Hole excitation: Time evolution of eigenenergies in the valence band (gray lines); the red line denotes the average energy of the excited hole. The right panels show the corresponding projected density of states (PDOS) calculated based on time \textit{t}=0 structure. We find that the states in the energy range that we are interested are dominated by PbI inorganic frame. The contribution of MA is out of this range. The PDOS plots also highlight the state ranges corresponding to different orbital characters. In panel (a), the energy zero point corresponds to the energy of the CBM at \textit{t}=0, while in panel (b), the energy zero point corresponds to the energy of the VBM at \textit{t}=0.}
\label{fig:dos-cooling}
\end{figure}

Since PbI$_2$ is usually excess in the precursor solution\cite{excessPbI2}, we focus on the PbI-terminated MAPbI$_3$ perovskite and analyze the structural distortion from three aspects:  
(1) The average deviations in lengths ($L$) and angles ($\theta$) of Pb-I vectors within PbI$_x$ polyhedra (surface layer (layer1) has PbI$_5$ polyhedra, second layer (layer2) has PbI$_6$ polyhedra) relative to ideal polyhedra (Equ.S1). Here, we only focus on the top two PbI$_x$ polyhedra layers since the bottom PbI$_x$ layer is fixed.
(2) The Pb–I–Pb angle between adjacent PbI$_x$ polyhedra (Equ.S2), which includes the Pb-I-Pb angles within the same layer ($\beta_{\text{layer1}}$, $\beta_{\text{layer2}}$) and between the layer1 and layer2 ($\beta_{\text{interlayer}}$).
(3) The rotation of the MA$^+$ molecule, represented by the autocorrelation function of the vector formed by C and N atoms, as defined in Equ.S3. Upon UV illumination with a wavelength of 280~nm (4.43~eV) corresponding to the maximum energy in the UVb region, either an electron or a hole is excited. For the electron excitation, the initial state of the excited electron is set at the valance band maximum (VBM)+4.43~eV, while for the hole excitation, the initial state of the excited hole is set at the conduction band minimum (CBM)--4.43~eV. Then, TDDFT-NOB is performed for each case to evolve the system with the excited electrons (or holes). Fig.\ref{fig:dos-cooling} shows the averaged energy of electrons and holes during the cooling process, both occurring on a timescale of about 1~ps. This agrees well with previously reported experimental and computational results\cite{expcooling,namdcooling}. Moreover, Fig.\ref{fig:dos-cooling} reveals that hole cooling is faster than electron cooling, which is consistent with earlier reports\cite{holefaster}. This is attributed to the much larger density of states (DOS) in the valence band than in the conduction band, which accelerates the hole relaxation\cite{nac}.

\begin{figure}[H]
\centering
\includegraphics[width=1\textwidth]{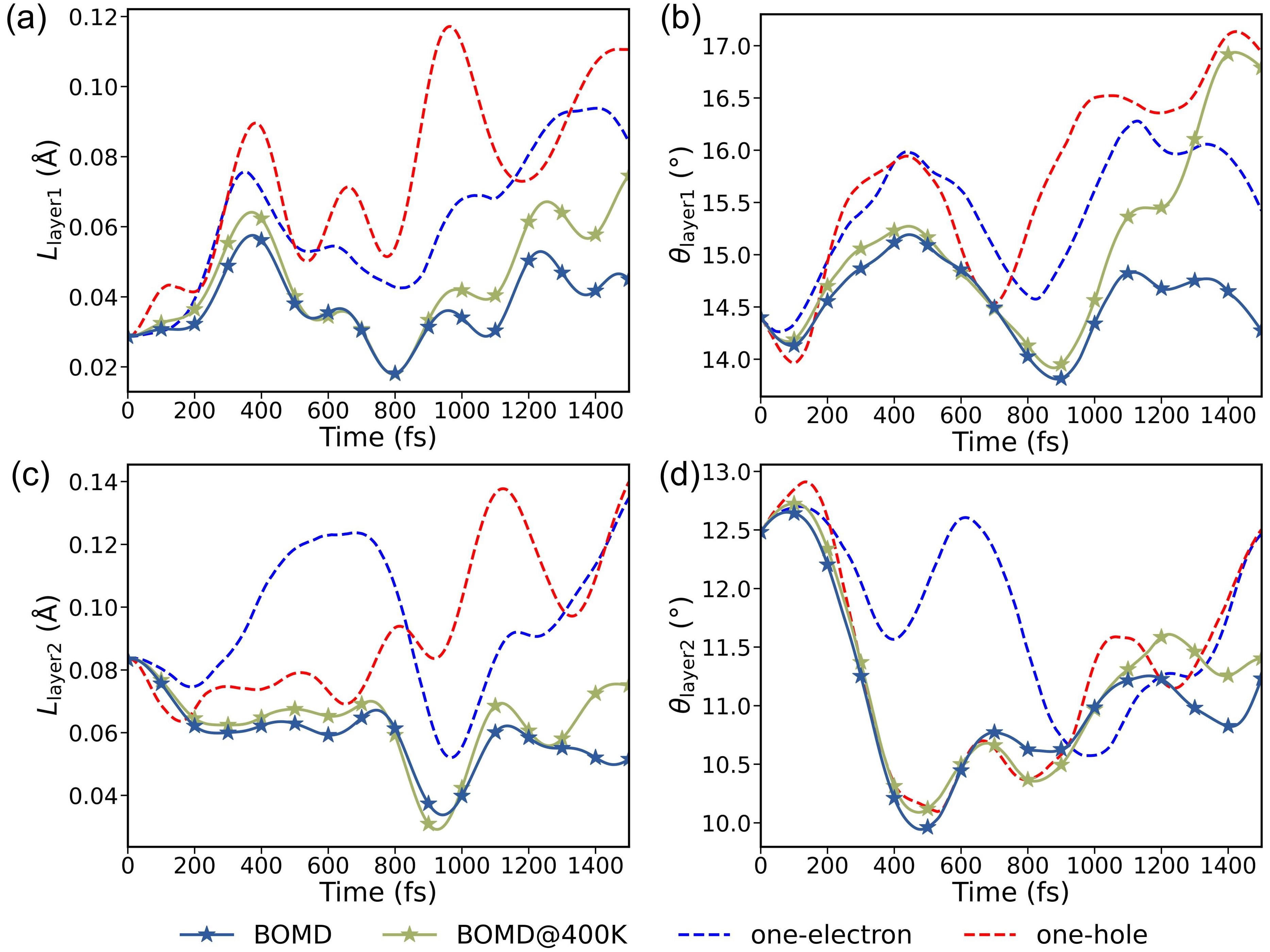}
\captionof{figure}{Time evolution of $L$ and $\theta$ (deviations to ideal PbI$_x$ polyhedral) for layer~1 and layer~2.}
\label{fig:nob-vector-mapbi3}
\end{figure}

To reveal the UV-induced distortion, we compare the TDDFT-NOB results with ground-state Born–Oppenheimer molecular dynamics (BOMD). To examine the effect of temperature increase during the cooling process on structural distortion, we also rescale the initial temperature to 400~K for the BOMD calculations. This is because Fig.~S2 shows that the system temperature increases to approximately 400~K at the end of the cooling process by the TDDFT-NOB simulation. Shown in Fig.\ref{fig:nob-vector-mapbi3} is the changes of $L$ and $\theta$ after photoexcitation of a hot carrier. Compared to the ground-state BOMD, UV-excited one electron or hole generally increases $L$ in both layer1 and layer2. For $\theta$, an increasing trend is observed in layer1, while in layer2, only the one-electron case shows an increase in the initial $\sim$800~fs; otherwise, it remains close to the ground-state values. These distortions in $L$ and $\theta$ suggest the deviations of the PbI$_x$ polyhedra from the ideal geometry. Moreover, as seen in Fig.\ref{fig:nob-vector-mapbi3}, the thermal effect is mainly evident after 1000~fs. The changes in $L$ and $\theta$ upon exciting a single electron or hole cannot be explained by the pure thermal effect, which is consistent with existing experimental reports\cite{heatvsUV}. To understand the causes of the variations in $L$ and $\theta$, we consider two mechanisms: (1) Carrier occupation of different orbitals: electrons in antibonding orbitals or holes in bonding orbitals will weaken the corresponding bonds and induce structural distortion; (2) Direct cooling induced distortion (DCID): During an NOB event, the wavefunction collapse transfers electronic energy to atoms, leading to change of atomic velocities and displacements. To distinguish these two mechanisms, we perform a special ground-state BOMD simulation in which the trajectory is starting with the same configuration to TDDFT-NOB, but the additional velocity from each NOB event is added to the respective atom of such BOMD simulation at the corresponding MD step of the NOB event. This procedure excludes the orbital occupation effects as all electrons are at the ground state, but only considers DCID effect. As shown in Fig.~S3, the changes in $L$ and $\theta$ for the hole excitation simulated by TDDFT-NOB and the special BOMD (denoted as one-hole-velocity-NOB) are very similar, while those for the electron excitation by TDDFT-NOB and the special BOMD (denoted as one-electron-velocity-NOB) differ significantly. This indicates that structural distortion in the single hole case is mainly driven by the DCID mechanism, whereas in the electron case, other factors contribute to the distortion. Next, we consider the impact of orbital occupation by carriers. Fig.\ref{fig:dos-cooling} shows the dominant orbital characters of MAPbI$_3$ across different energy regions. We performed constrained-BOMD simulations in which one electron or one hole is evenly distributed among states within each orbital-characterized energy region and their occupations remain unchanged during BOMD simulation. As shown in Fig.~S4, the most pronounced structural distortions occur when electrons occupy Pb-\textit{p} and I-\textit{p} antibonding states, or when holes occupy Pb-\textit{p} and I-\textit{p} bonding states. However, the different cooling time for the electrons and holes will play a role. Fig.\ref{fig:dos-cooling} reveals that the excited-state electron remains in the states with Pb-\textit{p} and I-\textit{p} antibonding orbital character for approximately 800~fs, while the hole occupies states with Pb-\textit{p} and I-\textit{p} bonding orbital character for only less than 200~fs. This long-time electron occupation in the antibonding states leads to a great structural distortion during the initial 800~fs, resulting in significant differences between the TDDFT-NOB and electron-velocity-NOB cases (Fig.~S3). In contrast, the short duration of hole occupation in the bonding states explains the small difference between the one-hole simulation by TDDFT-NOB and its special BOMD (one-hole-velocity-NOB).

\begin{figure}[H]
\centering
\includegraphics[width=1\textwidth]{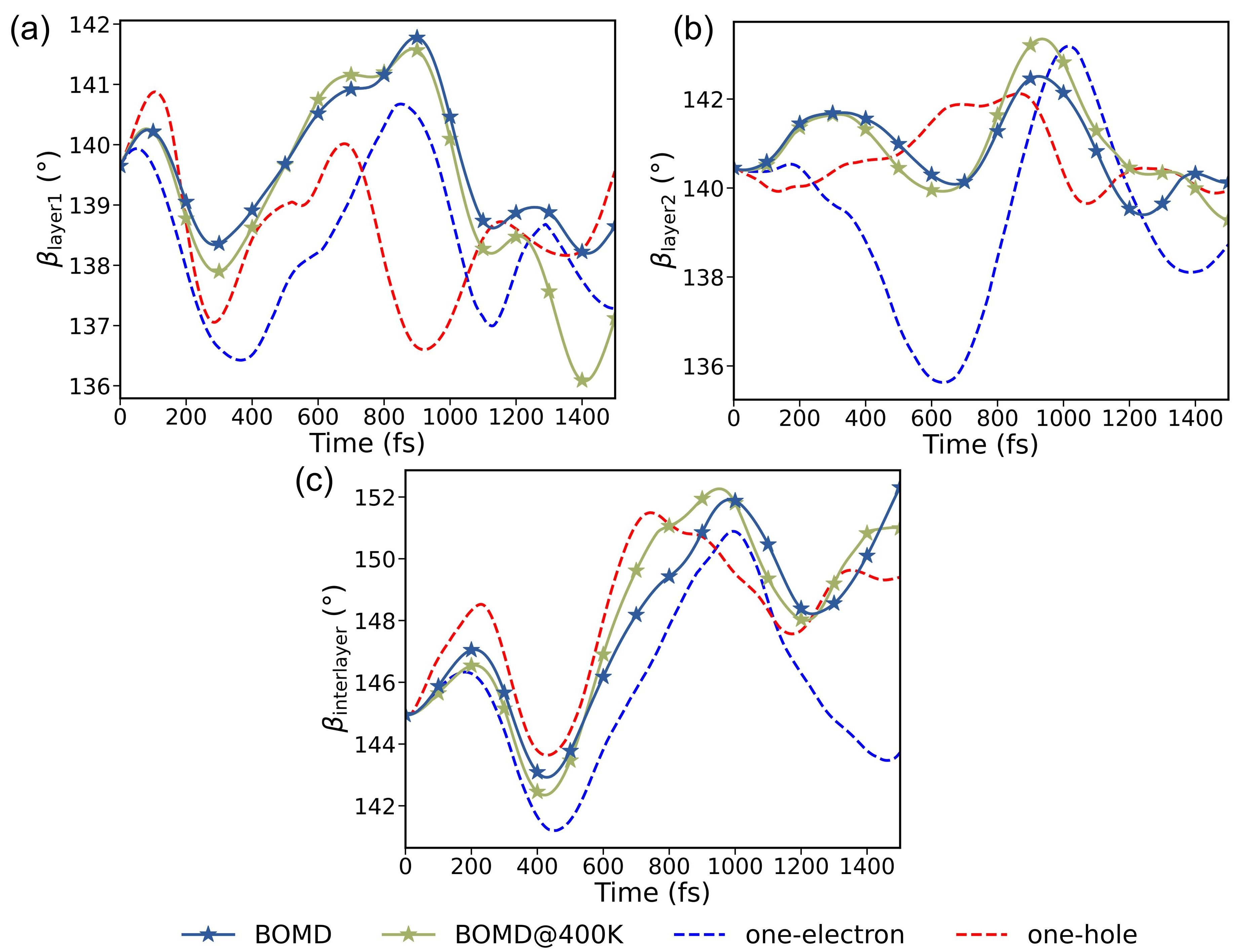}
\captionof{figure}{Time evolution of $\beta$ (in-plane Pb-I-Pb angles) for layer~1, layer~2, and the interlayer (out-of-plane Pb--I--Pb angle).}
\label{fig:nob-angle-mapbi3}
\end{figure}

In addition to polyhedral distortion, another important structural parameter is the Pb--I--Pb angle connecting two adjacent polyhedra. A decrease in this angle implies a high tendency to degradate into PbI$_2$. This is because such Pb--I--Pb angle in PbI$_2$ is approximately $90^\circ$ which is much smaller than the typical $160^\circ$--$180^\circ$ observed in tetragonal MAPbI$_3$. As shown in Fig.\ref{fig:nob-angle-mapbi3}, Pb-I-Pb angles experience structural distortion by the excited electron or hole to various degrees. Similar to $L$ and $\theta$, the uniform thermal effect has little impact on $\beta_{\text{layer1}}$, $\beta_{\text{layer2}}$ and $\beta_{\text{interlayer}}$. In order to distinguish the impact of electron's (or hole's) occupation and the DCID during the carrier cooling to the structural distortion, we have also investigated the special BOMD mentioned above by including the NOB-adjusted velocity (Fig.~S5) and the constrained BOMD (Fig.~S6). For all the three $\beta$ angles studied here, the structural deviations computed from the TDDFT-NOB of hole are always consistent to the calculations of the hole's special BOMD (denoted as one-hole-velocity-NOB, see Fig.~S5). It indicates that the hole plays its role mainly through the DCID mechanism. However, for electrons, by comparing Fig.~S5 and Fig.~S6, the induced deviations are mainly contributed by the electron occupation in states with Pb-\textit{p} and I-\textit{p} antibonding orbital character. The DCID contribution by electron is relatively small. Although the hole's occupation in states with Pb-\textit{p} and I-\textit{p} bonding orbital character may also introduce structural distortions (Fig.~S6), as we discussed for $L$ and $\theta$, the relatively short time occupation of these states by the hole reduces the overall effect of this mechanism.

\begin{figure}[H]
\centering
\includegraphics[width=0.8\textwidth]{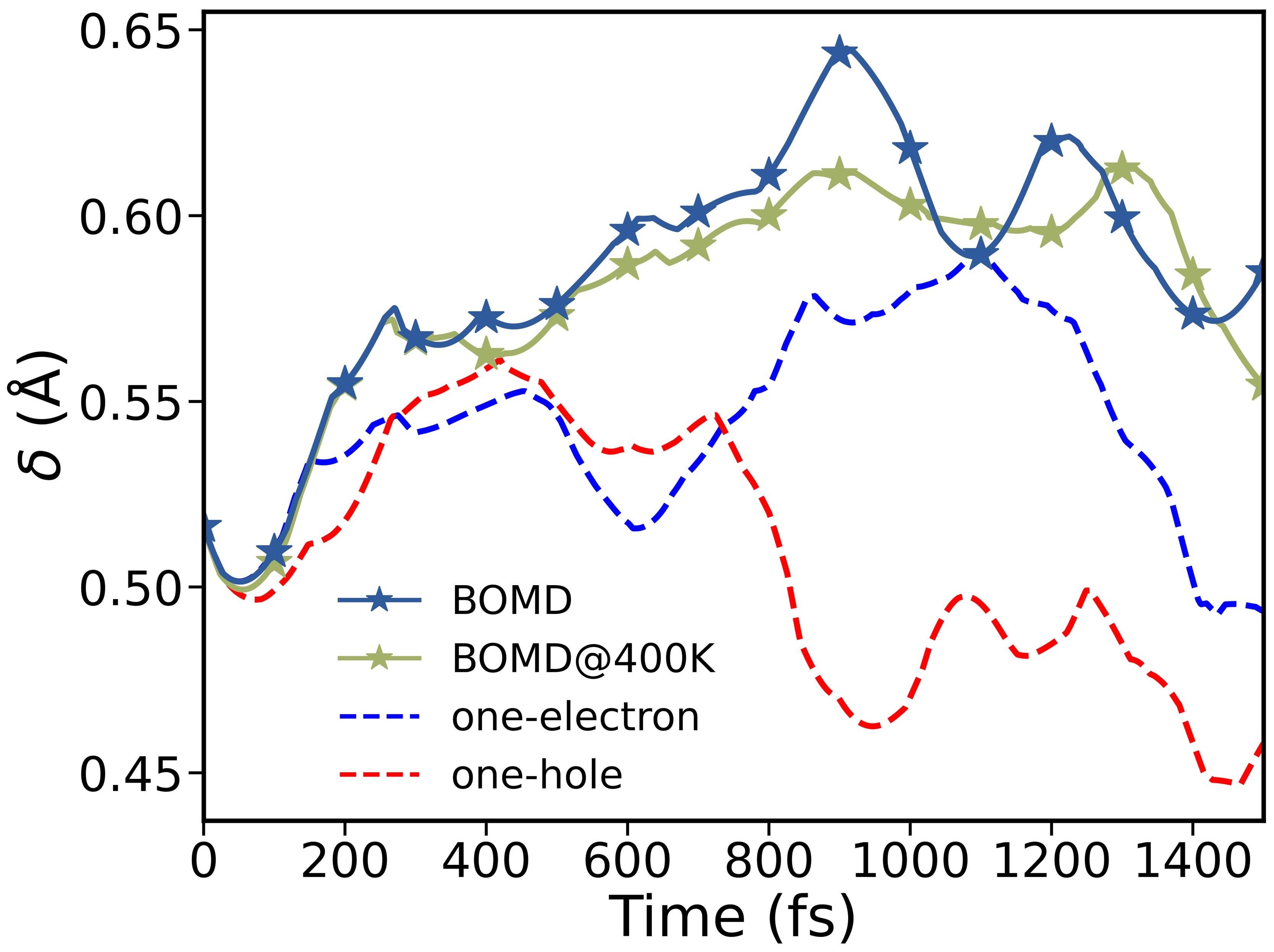}
\captionof{figure}{The evolution of $\delta$.}
\label{fig:nob-min}
\end{figure}

To directly quantify the formation of local PbI$_2$-type structure, we measure the mean deviations of the six lengths in the local Pb$_2$I$_2$ quadrilateral structures of MAPbI$_3$, denoted as $\delta$. A smaller $\delta$ indicates a more probable formation of local Pb$_2$I$_2$ quadrilateral structure similar to PbI$_2$ crystal (for details, see the Pb$_2$I$_2$ quadrilateral section in the Supporting Information). As seen in Fig.\ref{fig:nob-min}, $\delta$ decreases significantly in both the one-hole and one-electron cases compared to the ground-state BOMD, indicating a trend of forming PbI$_2$-type lattice. As shown in Fig.~S8 and Fig.~S9, we also measure $\delta$ for the special BOMD simulation and the constrained BOMD to illustrate the microscopic mechanisms for the electron and hole, respectively. They are consistent with the $\beta$ angle discussed above.

\begin{figure}[H]
\centering
\includegraphics[width=0.8\textwidth]{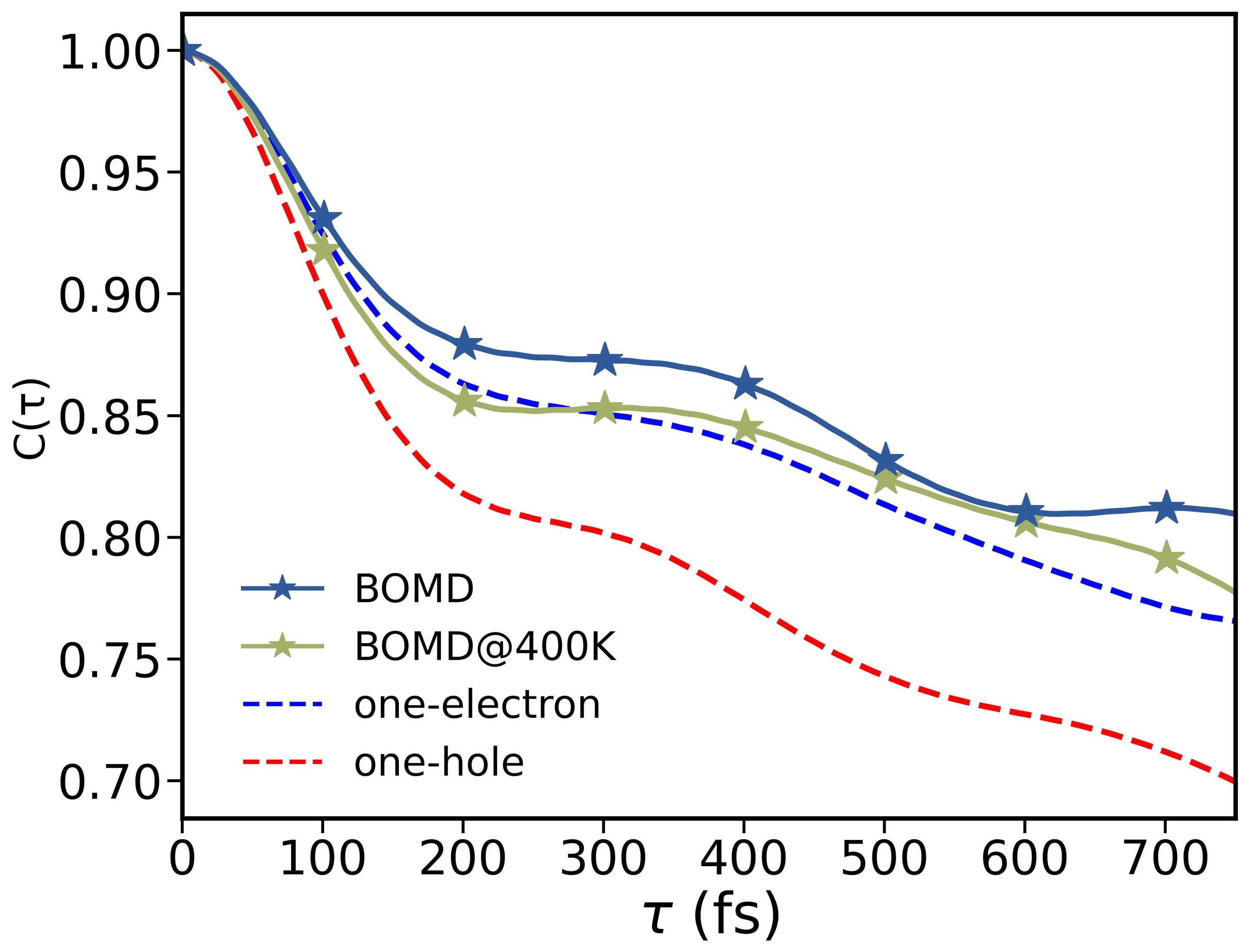}
\captionof{figure}{The autocorrelation function of MA$^+$ orientation as a function of time. To ensure sufficient sampling for averaging, $\tau$ is set to half of the total simulation time, i.e., 750~fs.}
\label{fig:nob-ma-mapbi3}
\end{figure}

Next, the behavior of MA$^+$ molecules is examined, focusing on their dissociation and rotation. This is because the dissociation of MA$^+$ molecules has been reported to trigger the decomposition of MAPbI$_3$, while MA$^+$ rotation facilitates halide ion migration by re-establishing hydrogen bonds \cite{degration-MA,migration-MA}. During the TDDFT-NOB simulation, we haven't observed the decomposition of MA$^+$ molecules. This is also shown by the maximum bond lengths within the MA$^+$ molecule (Table S1). This is mainly because the molecular orbitals are out of the energy range of UV light uesd here (Fig.\ref{fig:dos-cooling}).  Fig.\ref{fig:nob-ma-mapbi3} demonstrates that MA$^+$ rotation accelerates under both the electron or hole excitations. It is interesting to find that the rotation is significantly faster in the hole excitation than in the electron excitation. Analysis of the kinetic energy transferred to MA$^+$ via DCID (Fig.~S10) clearly shows a greater energy allocation in the hole case than that in the electron case, which explains the faster rotational speed. The temperature only slightly accelerates MA$^+$ rotation. Such observation is also consistent to the case of UV on the inorganic sublattice: For the hole excitation, the induced structural deviations is mainly through the DCID mechanism, where the phonon modes of both inorganic lattice and MA$^+$ molecules (this is explored more in the discussion below) are excited to distort the structure. However, for the electron excitation, its role is mostly through the occupation of states with Pb-\textit{p} and I-\textit{p} antibonding orbital character, which is less related to the MA$^+$ molecules.

With the microscopic mechanisms unveiled, we want to propose a way to mitigate the UV induced structural distortions. For MAPbI$_3$, the high electron mobility allows excited electrons to easily migrate to the electron transport layer\cite{mobility}, and the effect of electron occupying states with Pb-\textit{p} and I-\textit{p} antibonding states may not be significant. Thus, finding a way to suppress the DCID contribution will be crucial. Here, we design a molecule, 2,3-Butanedione dioxime (BDO), with two oxime groups to stabilize the lattice through a multidentate coordination with surface Pb and I. On one hand, the strong binding of this molecule can help to saturate the surface Pb atoms and stablize the lattice. On the other hand, we expect the strong binding can also mitigate part of the DCID effect and reduce the distortion to MAPbI$_3$ lattice. With this molecule adsorbed to the surface, we re-perform the TDDFT calculation and analyze all the metrics aforementioned. For easy comparison, each metric is obtained by subtracting the corresponding ground-state BOMD result and denoted with a ``$\Delta$'' symbol. Values closed to zero indicates that they conform to the BOMD at equilibrium. After the incorporation of BDO, for the hole case, the increasing trends of $L_\mathrm{layer1}$, $\theta_\mathrm{layer1}$ and $L_\mathrm{layer2}$ (Fig.~S11), $\beta_{\text{layer1}}$, $\beta_{\text{layer2}}$ and $\beta_{\text{interlayer}}$ (Fig.~S13), and $\delta$ (Fig.~S14) are all suppressed significantly. For the one-electron case, the benefit is less manifest compared to the hole case as we expected. This is because the incorporation of BDO does not alter the Pb-\textit{p} and I-\textit{p} antibonding characters of the MAPbI$_3$ conduction band (Fig.~S12). Furthermore, we investigate changes in MA$^+$ rotational dynamics upon BDO adsorption. As shown in Fig.~S15, under one-hole excitation, the rotational speed of MA$^+$ molecules is consistently slower after BDO adsorption.

\begin{figure}[H]
\centering
\includegraphics[width=1\textwidth]{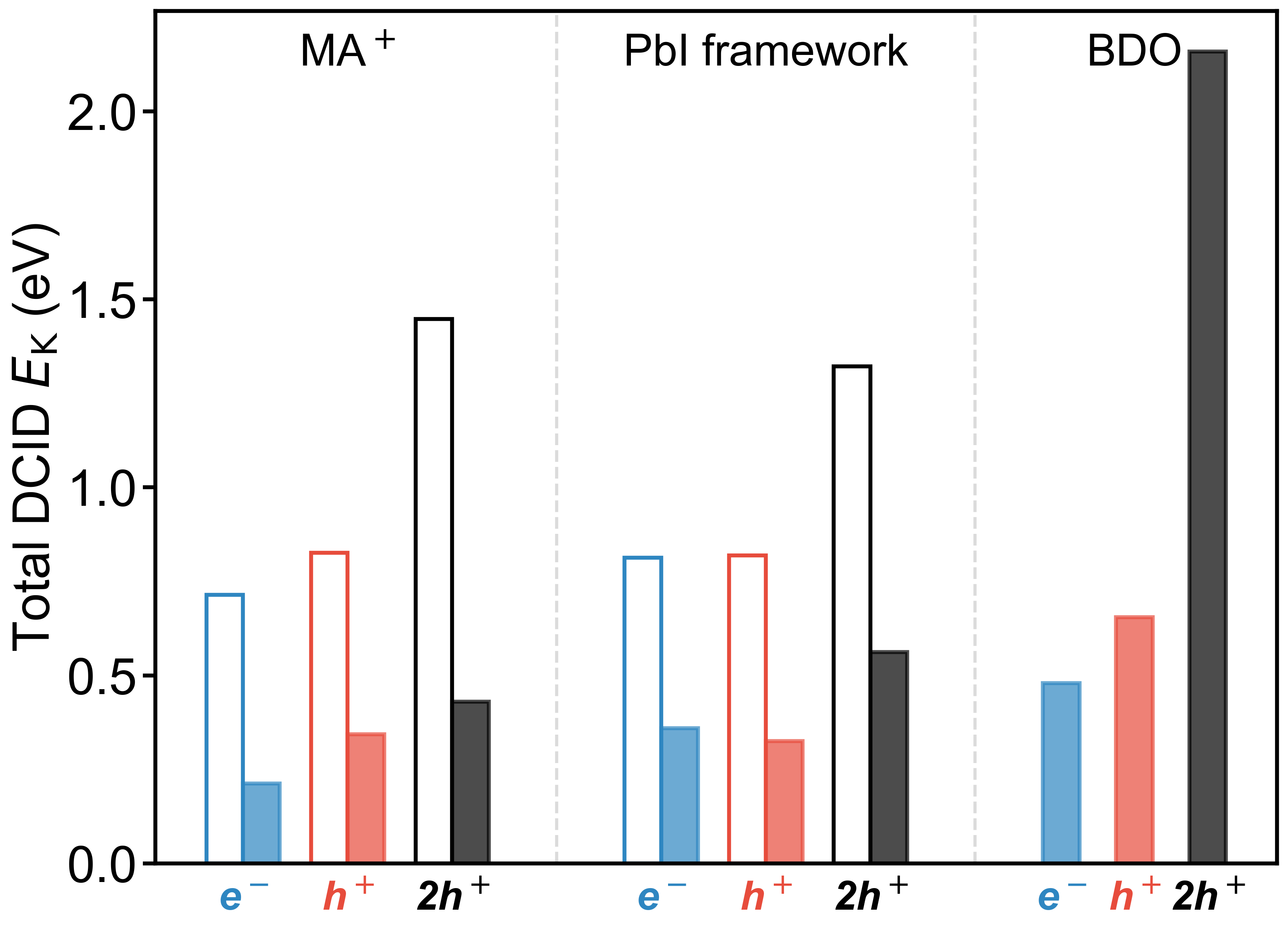}
\captionof{figure}{The total kinetic energy induced by the NOB event (\ensuremath{E_{k,\text{tot}}}) for different components: the organic part (MA$^+$), the inorganic part (Pb and I), and the BDO passivator. Bars with color filling represent the system with BDO.}
\label{fig:ek}
\end{figure}

To understand how BDO molecules influence the structural changes in MAPbI$_3$, we analyzed the total kinetic energy (\ensuremath{E_{k,\text{tot}}}) transferred to the lattice by NOB events to different components of the system with and without BDO adsorption under various excitation conditions. As shown in Fig.\ref{fig:ek}, under both one-electron or one-hole excitation, the \ensuremath{E_{k,\text{tot}}} acquired by the inorganic framework (PbI) and the MA$^+$ molecules is significantly reduced when BDO molecules are adsorbed on the MAPbI$_3$ surface, while the BDO molecules receive considerable kinetic energy. This behavior can be understood from the phonon normal-mode analysis. We compute the normal modes of the system by diagonalizing the dynamical matrix under the harmonic approximation. The DCID induced atomic velocity change is further decomposed onto these normal modes (Equ.S9) to obtain the corresponding normal-mode coordinates. These coordinates can directly reveal the excited phonon modes by DCID processes. Fig.~S16 shows the accumulated decompositions to all normal modes throughout the whole trajectory. For bare MAPbI$_3$ case, these decompositions are mostly distributed on PbI lattice ($\leq 5 \, \text{THz}$) and MA$^+$ molecules (spikes $\geq 5 \, \text{THz}$). It is also interesting to note that the phonon modes around 5 THz just correspond to the MA$^+$ rotation. The large distribution of normal-mode coordinates can also explain the aforementioned increased MA$^+$ rotational speed. However, the phonon-mode gaps (e.g. from 6 Thz to 30 THz) aggravate the DCID induced energy to PbI lattice and MA$^+$ and distort the MAPbI$_3$ lattice. By passivating BDO molecule, we find that such molecule provides many phonon modes within the previous phonon gaps, allowing the multi-channels to dissipate the energy to BDO instead of PbI inorganic lattice. In the end, it will enhance the UV stability.

\begin{figure}[H]
\centering
\includegraphics[width=1\textwidth]{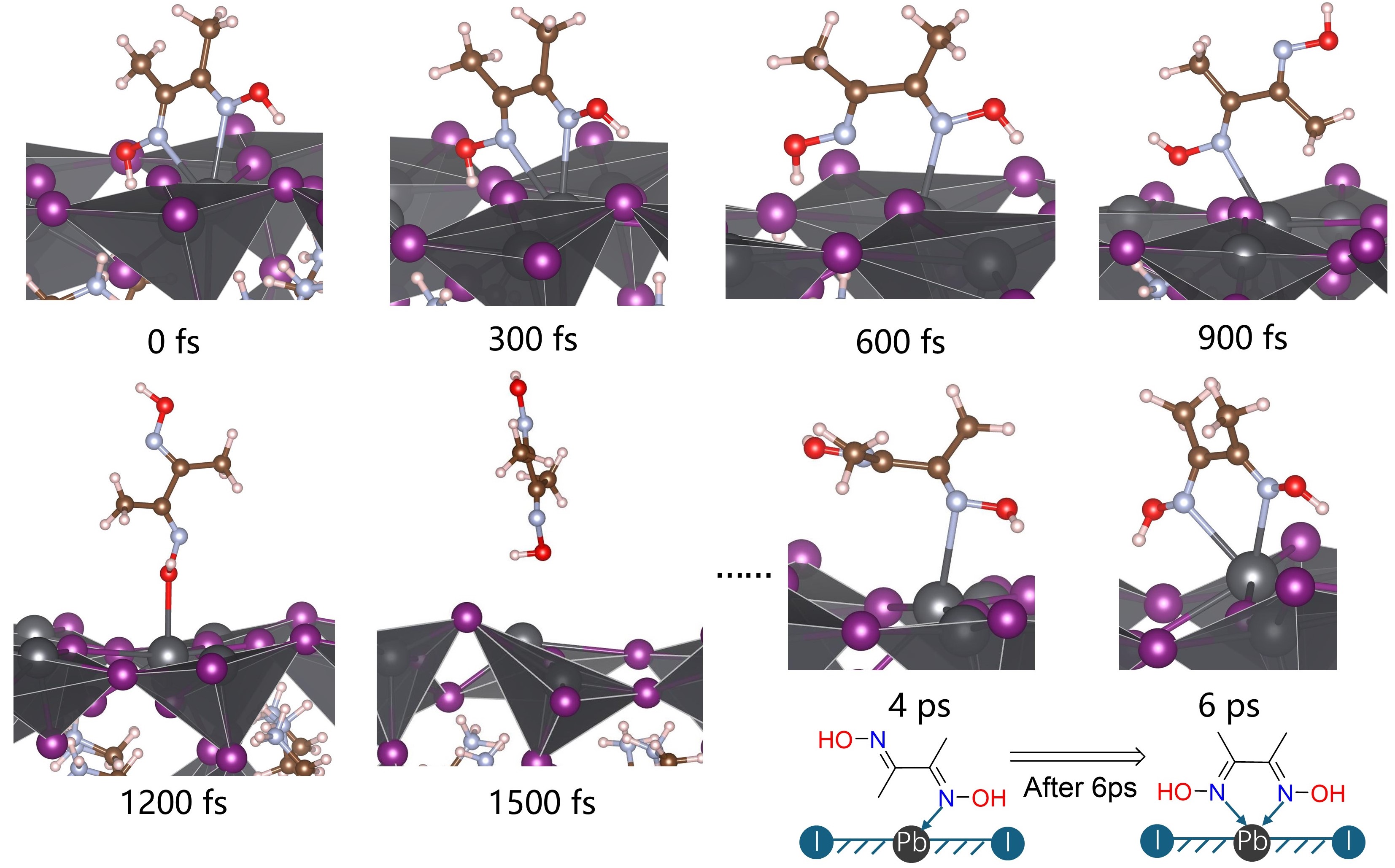}
\captionof{figure}{Trajectory snapshots of BDO molecular conformational change and desorption during the 1500 fs TDDFT-NOB two-hole simulation. The 4ps and 6ps snapshots are from a 10 ps ground-state BOMD simulation, initialized with atomic velocities and positions at 1500 fs of the TDDFT-NOB two-hole simulation.}
\label{fig:nob-snapshot}
\end{figure}

The benefit of BDO to enhance UV-instability is also tested for multiple excitations, where more than one excited electrons or holes are excited\cite{multiple-excitation}. Since the hole's mobility is much smaller than the electron in MAPbI$_3$\cite{mobility}, the holes are more easily to accumulate in a small region\cite{hole-trap} (for example near a defect). For this reason, we focus on the case of exciting two holes simultaneously. In this case, the effect of BDO is more pronounced, as illustrated in Fig.~\ref{fig:ek}. By exciting two holes, BDO can absorb the substantial DCID energy from carrier cooling process and reduce such energy allocated to PbI or MA$^+$ significantly. All the metrics including Pb-I bond lengths ($L$) and angles ($\theta$) (Fig.~S17), Pb-I-Pb angles ($\beta$) (Fig.~S18), local PbI$_2$-type structure ($\delta$) (Fig.~S19), and rotation of MA$^+$ (Fig.~S20) are suppressed effectively by BDO adsorption. Particularly, as shown in the structural snapshots at 1.5 ps (end of carrier cooling), no Pb$_2$I$_2$ structures form when BDO is added. The structural distortion of MAPbI$_3$ with BDO is significantly smaller than that without BDO (Fig.~S21). Furthermore, we want to identify an intriguing phenomenon. When BDO acquires sufficient kinetic energy—such as two‑hole excitation—the molecule can undergo a conformational change (see Fig.\ref{fig:nob-snapshot}): the two oxime groups switch from a favorable "\textit{cis}-type" adsorption configuration to an unfavorable "\textit{trans}-type" conformation, followed by desorption from the MAPbI$_3$ surface. While such UV‑induced molecular desorption has been reported experimentally before\cite{uvdesoprtion}. In our study, The BDO molecule plays its role by reverting to the favorable conformation and readsorbing onto the MAPbI$_3$ surface, occurring 6~ps after the cooling process ends (Fig.\ref{fig:nob-snapshot}). Such conformation change and readsorption process help the molecule to dissipate the acquired energy without breaking apart. The benefit of BDO originates from its strong and multi-coordinating bindings with MAPbI$_3$ surface. It enables this molecule to collect the enrgy generated during the hot carrier cooling process and reduces the contribution of DCID. As a comparison, we select a different molecule but with a weaker binding than BDO to demonstrate. A commonly used SAM molecule in PSCs, specifically 2,3,4,5,6-Pentafluorobenzylphosphonic acid (PFPA) is tested.\cite{5F-1,5F-2}. As shown in Fig.\ref{fig:ek}, since the effect of BDO is more pronounced under two-hole conditions, we illustrate the comparison between PFPA and BDO under two-hole conditions. We also perform TDDFT calculations of one-hole excitation for PFPA adsoprtion and their comparison to the one-hole BDO case shows similar trends to the two-hole case. From Figs.~S22--S24, PFPA does not effectively suppress MAPbI$_3$ structural distortion compared to BDO. Further analysis of the NOB-induced kinetic energy obtained by BDO and PFPA (Fig.~S25) explains that the kinetic energy acquired with PFPA is much lower than with BDO. This is further consistent to the phonon decomposition result. We analyze the accumulated $P_\nu$ obtained from different passivators' phonon modes (one-hole and one-electron cases, Fig.~S16). In the phonon mode gap region of pristine MAPbI$_3$ (6~THz to 30~THz), the accumulated $P_\nu$ on BDO in the hole case is significantly larger than that on PFPA.

\begin{figure}[H]
\centering
\includegraphics[width=1\textwidth]{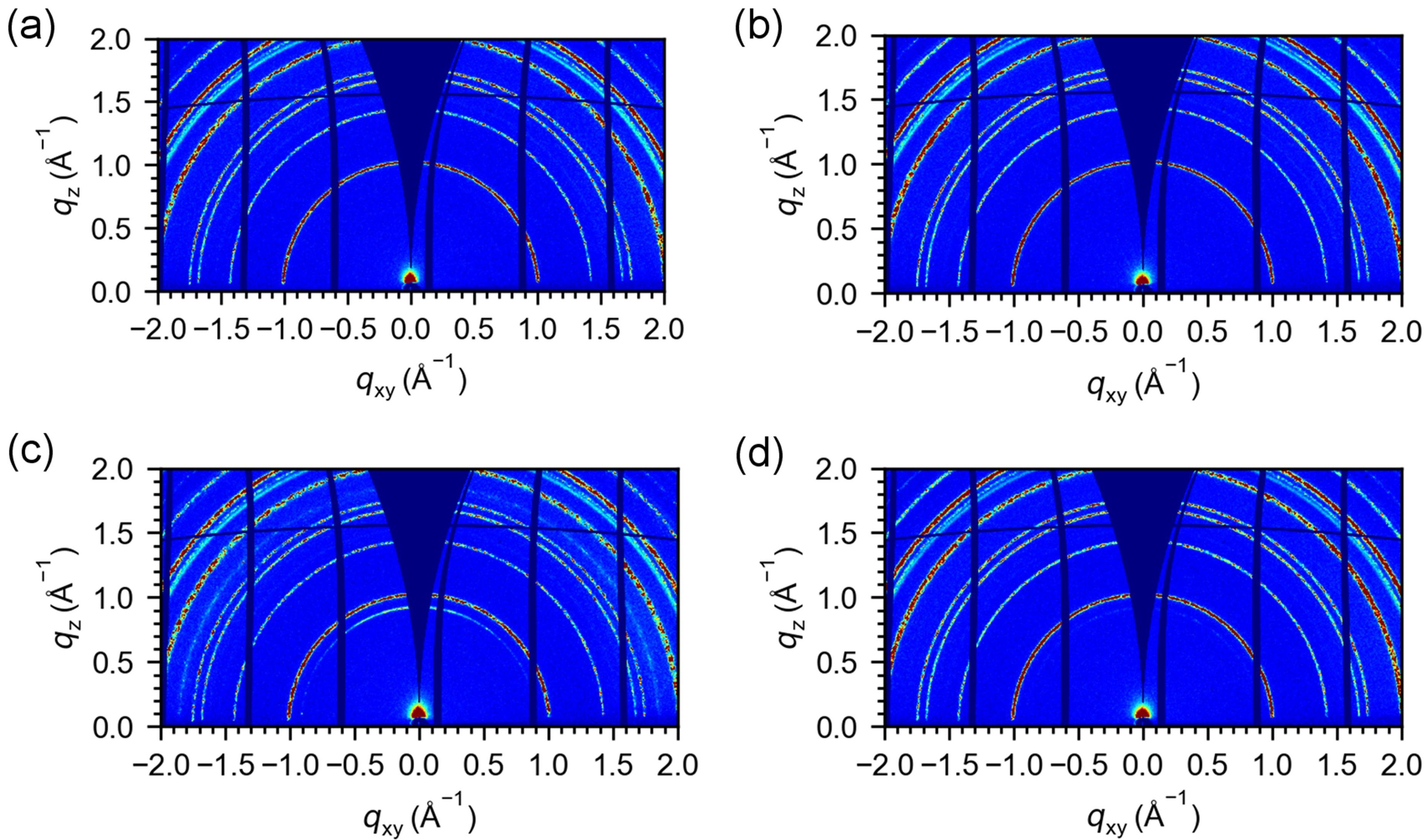}
\renewcommand{\thefigure}{\arabic{figure}}
\caption{GIWAXS patterns: (a) and (b) correspond to pristine MAPbI$_3$ and MAPbI$_3$ with BDO, respectively; (c) and (d) correspond to the same films after 24~h UVB (310~nm) exposure at 50~$^\circ$C.}
\label{fig:GIWAXS}
\end{figure}

\begin{figure}[H]
\centering
\includegraphics[width=1\textwidth]{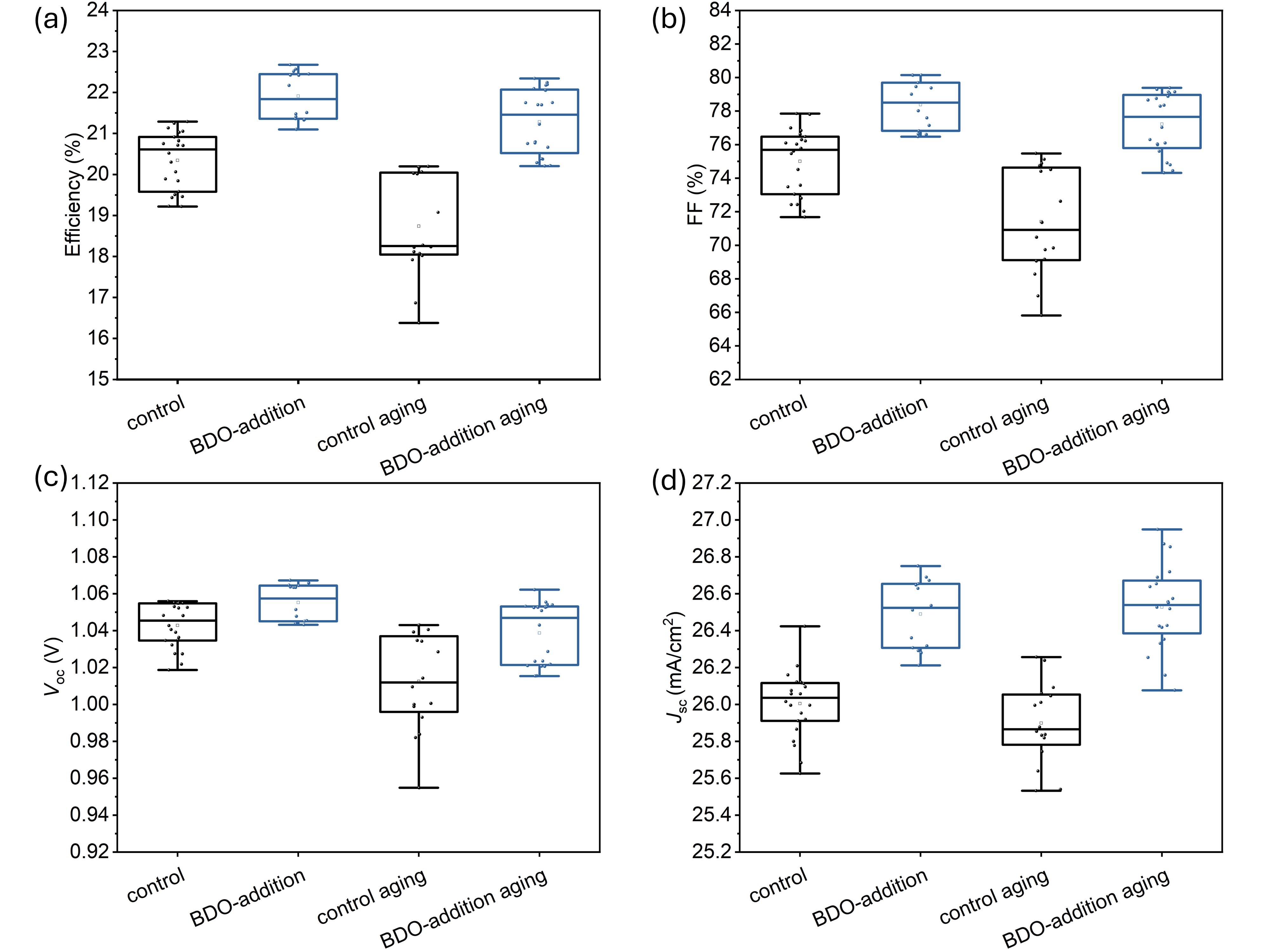}
\renewcommand{\thefigure}{\arabic{figure}}
\caption{(a) Device efficiency, (b) fill factor (FF), (c) open-circuit voltage ($V_\mathrm{oc}$), and (d) short-circuit current density ($J_\mathrm{sc}$) of PSCs. “Control” refers to PSCs without BDO, while “BDO-addition” indicate PSCs with 4~mg/mL BDO molecules. “Aging” represents devices exposed to 310~nm UV light at 50~$^\circ$C for 12~hours.}
\label{fig:device}
\end{figure}

Finally, we experimentally validate the UV-protective effect of BDO molecules. As shown in Fig.\ref{fig:GIWAXS}, the GIWAXS patterns for both pristine MAPbI$_3$ (a) and MAPbI$_3$ with BDO (b) exhibit the typical diffraction features of MAPbI$_3$, with prominent peaks corresponding to the perovskite phase. After exposure to 310~nm UV light at 50~$^\circ$C for 24 hours, significant changes are observed in the GIWAXS patterns of pristine MAPbI$_3$ (c). A strong PbI$_2$ diffraction signal emerges, especially in the low-$q$ region, indicating the formation of a secondary PbI$_2$ phase near the surface as a result of UV degradation. In contrast, MAPbI$_3$ with BDO (d) remains largely unaffected by the UV exposure, with no significant increase in the PbI$_2$ signal. To rule out thermal effect of degradation, we also hold MAPbI$_3$ at 50~$^\circ$C for 24~hours without UV exposure (Fig.~S26), which showed no PbI$_2$ formation—indicating that BDO specifically protects against UV-induced degradation. Fig.\ref{fig:device} shows the device performance with and without BDO addition, both before and after aging. The boxplots demonstrate that BDO-addition PSCs exhibit significantly improved stability under UV exposure compared to the control devices. In particular, the key performance indicators—efficiency, fill factor (FF), open-circuit voltage ($V_\mathrm{oc}$), and short-circuit current density ($J_\mathrm{sc}$)—remain more stable in the BDO-addition devices after 12 hours of exposure to 310 nm UV light at 50$^\circ$C. For the control PSCs without BDO, a noticeable decrease in efficiency, FF, and $V_\mathrm{oc}$ is observed after aging, indicating significant UV-induced damages. This analysis indicates that the presence of BDO significantly enhances the UV stability of MAPbI$_3$-based PSCs, making them more resilient to prolonged UV exposure compared to the control devices.

In conclusion, we have conducted a comprehensive investigation into the structural dynamics of PbI-terminated MAPbI$_3$ under 280nm UV excitation and elucidate the role of passivation molecules in enhancing its UV stability. Through TDDFT-NOB simulations, we identify that the uniform increase of temperature after UV illumination is not the main reason to triger the structural distortion. Instead, under the electron excitation, the distortion of the inorganic framework (PbI) is primarily driven by the electron occupation of Pb-\textit{p} and I-\textit{p} antibonding states, whereas in the hole case, it is mainly governed by the direct cooling induced distortion. We also find that UV accelerates the rotation of MA$^+$ molecules. To improve UV sbaility of MAPbI$_3$, we introduce BDO as a surface passivant. We find that BDO can facilitate the transfer of kinetic energy into BDO-specific vibrational modes, reducing the energy acquired by Pb or I atoms, thereby suppressing their displacement. Interestingly, under two holes excitation, BDO molecules undergo transient conformational changes and readsorb during thermal relaxation, maintaining their UV-protective function. We further compare the BDO molecule with the commonly used SAM molecule PFPA and find that BDO dissipates the potential energy from carrier cooling more effectively. Finally, experimental validation confirms that BDO-addition significantly improves the UV stability of both MAPbI$_3$ films and MAPbI$_3$-based PSCs, with suppressed PbI$_2$ formation and enhanced operational stability under prolonged UV irradiation. This work reveals the microscopic mechanism of UV degradation in MAPbI$_3$ and clarifies how passivating molecules enhance UV stability, providing a deep insight into the UV stability of perovskites to guide the design of improved strategies.

\section{Computational details}
\indent

All calculations are performed using the PWmat software package\cite{pwmat}. Optimized norm conserving pseudopotentials describe the electron-ion interactions\cite{psp}. The Perdew-Burke-Ernzerhof (PBE) exchange-correlation functional is used with the D3 dispersion correction\cite{gga-pbe,d3}. Table~S1 shows that the lattice parameters calculated with PBE-D3 are in best agreement with the experimental values, and the calculated band gap (1.61~eV) is also consistent with the experimental value of 1.58~eV\cite{latticepara}. For TDDFT, the time step is 0.1~fs, while for BOMD it is 1~fs. The plane-wave cutoff energy for the wavefunction is 50~Ryd, and the SCF iterations use the charge density as the convergence criterion with a threshold of $5\times10^{-5}\ e/\text{Bohr}^3$. The $\Gamma$-point is used for both TDDFT and BOMD calculations. The Fermi-Dirac $k_BT$ smearing energy is set to 0.1~eV. In TDDFT-NOB simulations, the cut-off entropy for branching is 0.5 and the decoherence time is 15~fs. During TDDFT calculations, the transition degree of freedom is first determined, and branching is allowed if it has sufficient kinetic energy to compensate the potential energy decrease. Phonon calculations are performed using the Phonopy package with a displacement of 0.04~\text{Å}, considering only the $\Gamma$-point\cite{phonopy1,phononpy2}. For phonon calculations, the SCF charge density convergence criterion is $7\times10^{-7}\ e/\text{Bohr}^3$, and the force convergence criterion is 0.002 eV/\AA.

To obtain the 300 K equilibrium structure, a 2 ps Langevin NVT BOMD simulation is performed. Structures and velocities from the equilibrium period (after 1.5 ps) close to 300 K are used as the initial information for a 10 ps Verlet NVE BOMD simulation. Snapshots at 2, 4, 6, and 8 ps from such NVE simulation are used as the initial coordinates and velocities for the TDDFT-NOB simulations. Different random seeds are applied for each calculation to ensure the stochasticity. If the excited hot electron (or hole) cools to the CBM (or VBM) before 1500 fs (all TDDFT-NOB calculations run for at least 1 ps), we perform constrained BOMD calculations by placing one electron in the CBM (or one hole in the VBM). Similarly, using these four snapshots as the initial configurations, one electron (or one hole) is distributed across states belonging to each orbital characterized region (Fig.\ref{fig:dos-cooling}), and a 1500 fs constrained BOMD calculation is performed. This is to analyze the effect of carrier occupying states with different orbital characters to the structural distortions. Additionally, using the same four snapshots, a special BOMD simulation is performed: when a BOMD step corresponds to a TDDFT-NOB time step at which a NOB event occurs, the BOMD is paused, and the extra velocity from the NOB event is added to the atoms at this time step. The adjusted velocity is then used to continue the BOMD calculation until it reaches 1500 fs. All structural distortion parameters shown in this paper are averaged over the four trajectories.

\section{Experimental details}
\section{Fabrication of MAPbI$_3$ Perovskite Films}
\indent

MAPbI$_3$ microcrystals are synthesized via an aqueous method as previously reported \cite{lwz}. To prepare the wide-bandgap perovskite precursor, a 1.5~M solution is obtained by dissolving MAPbI$_3$ microcrystals with or without 10~mg of BDO in a mixed solvent of N,N-dimethylformamide (DMF) and dimethyl sulfoxide (DMSO) at a 3:1 volume ratio. The solution is stirred at 60~$^\circ$C for 1~h and filtered through 0.22~µm polytetrafluoroethylene (PTFE) filters before deposition. Substrates are treated with ultraviolet–ozone for 15~min to remove organic contaminants and enhance surface wettability. The 4PADCB solution (0.5~mg~mL$^{-1}$ in ethanol) is spin-coated onto the substrates at 5,000~rpm for 30~s and thermally annealed at 100~$^\circ$C for 10~min. The perovskite films are fabricated using a two-step spin-coating process: first at 3,000~rpm for 10~s (acceleration: 1,500~rpm~s$^{-1}$), followed by 5,000~rpm for 40~s (acceleration: 3,000~rpm~s$^{-1}$). During the second step, 200~µL of anisole is dynamically dripped onto the spinning substrate 10~s before the end of the process. The films are annealed at 100~$^\circ$C for 10~min on a hotplate. The perovskite films are then transferred to a hotplate at 50~$^\circ$C and exposed to ultraviolet irradiation using a 310~nm lamp with 12~W power for 24~h. The lamp is kept within 1~cm of the samples to ensure effective ultraviolet aging.

\section{Perovskite Solar Cells Fabrication}
\indent

FTO glass substrates are cleaned sequentially with detergent, deionized water, and isopropanol, and then treated with UV/O$_3$ for 20~min. For the hole transport layer, 4PADCB (0.5~mg/mL in ethanol) is spin-coated at 3000~rpm for 30~s and annealed at 110~$^\circ$C for 10~min. The perovskite precursor solution (Cs$_{0.05}$FA$_{0.95}$PbI$_3$, 1.6~M) is prepared by dissolving CsI, FAI, and PbI$_2$ in a DMF/DMSO mixed solvent (4:1 v/v) with 4~mg/mL BDO as an additive. A 60~$\mu$L aliquot of the precursor solution is dropped onto pre-annealed FTO substrates and spin-coated at 1000~rpm for 15~s and 5000~rpm for 35~s. During the last 10~s, 150~$\mu$L of CB is dropped as an antisolvent. The films are then annealed at 110~$^\circ$C for 10~min. C60 (28~nm, 0.04~nm/s), BCP (3~nm, 0.02~nm/s), and Ag electrodes (100~nm, 0.1~nm/s) are thermally evaporated under a vacuum of 5~$\times$~10$^{-6}$~Torr. All processes are conducted inside a nitrogen-filled glovebox with O$_2$ and H$_2$O levels maintained below 0.1~ppm.

\begin{acknowledgement}

This work is supported by the National Natural Science Foundation of China (62305215) and ShanghaiTech AI4S initiative SHTAI4S202404. The computational support is provided by the HPC platform of ShanghaiTech University.  
\end{acknowledgement}

\bibliography{ref}

\end{document}


\newpage

\begin{figure}[H]
\centering
\includegraphics[width=0.4\textwidth]{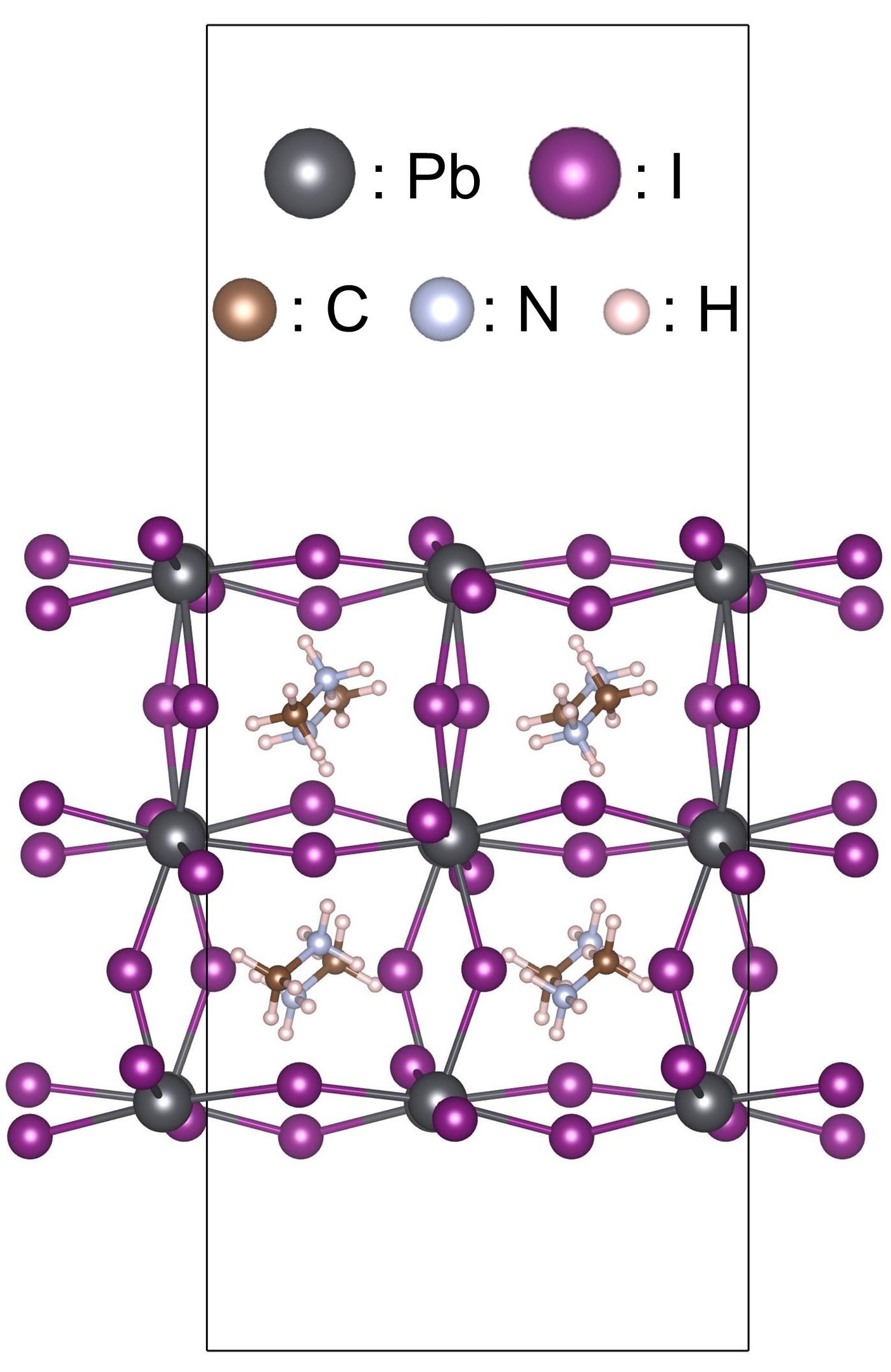}
\renewcommand{\thefigure}{S\arabic{figure}}
\caption{The PbI-terminated MAPbI$_3$ structure used for simulation.}
\label{fig:model}
\end{figure}

Starting from the tetragonal MAPbI$_3$ unit cell\cite{tetragonal,cif}, we cleave along (001) plane to bulid the PbI-terminated surface and construct a $\sqrt{2}\times\sqrt{2}$ supercell with a 17~\AA\ vacuum length and five-layer thickness (Fig.\ref{fig:model}). 

\renewcommand{\thetable}{S\arabic{table}}
\begin{table}[H]
\centering
\renewcommand{\arraystretch}{1.3}  
\begin{tabular}{lccc}
\hline
 \text{Method} & \text{a (Å)} & \text{b (Å)} & \text{c (Å)} \\
\hline
PBE       & 9.07 & 8.97 & 13.09 \\
PBE+D3    & 8.79 & 8.81 & 12.97 \\
Experiment\cite{latticepara} & 8.67 & 8.67 & 12.86 \\
\hline
\end{tabular}
\caption{Lattice parameters from different methods and experiment.}
\label{tab:lattice_parameters}
\end{table}

The perfect polyhedra is constructed with the Pb--I bond lengths (i.e., 3.338~\AA) taken as the average bond lengths over the 10~ps ground-state BOMD. In the perfect polyhedra, I--Pb--I angles are only 90$^\circ$ or 180$^\circ$. To analyze structural deviations from the perfect polyhedra, each polyhedron was aligned to its corresponding perfect polyhedron using the quaternion-based characteristic polynomial (QCP) algorithm\cite{qcp}. The aligned coordinates were then used to compute $L$ and $\theta$.
\begin{equation}
\begin{aligned}
L(t) &= \frac{1}{N_{\text{traj}}} \sum_{j=1}^{N_{\text{traj}}} \left[ \frac{1}{N_v} \sum_{i=1}^{N_v} \left( \left|\mathbf{r}_i(t)\right| - \left|\mathbf{r}_i^0\right| \right) \right]_{\text{traj}-j} \\
\text{and} \quad \theta(t) &= \frac{1}{N_{\text{traj}}} \sum_{j=1}^{N_{\text{traj}}}\left[ \frac{1}{N_v} \sum_{i=1}^{N_v}  \arccos \left( \frac{\mathbf{r}_i(t) \cdot \mathbf{r}_i^0}{\left|\mathbf{r}_i(t)\right| \left|\mathbf{r}_i^0\right|} \right) \cdot \frac{180}{\pi} \right]_{\text{traj}-j}
\end{aligned}
\end{equation}

\noindent
$L(t)$ and $\theta(t)$ represent the deviations in lengths and angles of Pb--I vectors relative to those in perfect polyhedra at time $t$. $N_v$ is the total number of Pb--I vectors in one structure, and $N_{\text{traj}}$ is the number of trajectories. $\mathbf{r}_i(t)$ and $\mathbf{r}_i^0$ represent the Pb--I vector in the polyhedra and the corresponding vector in the ideal polyhedra, respectively. $|\cdot|$ indicates the length of vectors. To compute these quantities, we sum over all vectors at each time step and then average over $N_{\text{traj}}$ trajectories.

\begin{equation}
\beta(t) = \frac{1}{N_{\text{traj}}} \sum_{j=1}^{N_{\text{traj}}} \left( \frac{1}{N_{\text{angle}}} \sum_{i=1}^{N_{\text{angle}}} \beta_j(t) \right)_{\text{traj}-j}
\end{equation}
\noindent
$\beta(t)$ represents the average Pb--I--Pb angles at time $t$. $N_{\text{angle}}$ is the total number of Pb--I--Pb angles and $\beta_j(t)$ is the $j$-th Pb--I--Pb angle.

The orientational autocorrelation function $C(\tau)$ of MA$^+$ molecules is defined as:
\begin{equation}
C(\tau) = \frac{1}{N_{\text{traj}} N_t N_p} \sum_{j=1}^{N_{\text{traj}}} 
\Bigg( \sum_{t=1}^{N_t} \sum_{i=1}^{N_p} 
\frac{\mathbf{r}_i(t) \cdot \mathbf{r}_i(t+\tau)}
{\left|\mathbf{r}_i(t)\right| \left|\mathbf{r}_i(t+\tau)\right|} 
\Bigg)
\end{equation}

\noindent
where $N_t$ is the number of time steps available for averaging (depending on the total number of time steps and $\tau$), $N_p$ is the number of C--N pairs per step, and $\mathbf{r}_i(t)$ is the vector from C to N atom of the $i$-th pair at time $t$.

\begin{equation}
\overline{E}(t) = \sum_{i=1}^{N} \epsilon_i(t) \, g_i(t), \quad 
g_i(t) = \sum_{j} \left| \langle \phi_i(t) | \psi_j(t) \rangle \right|^2
\end{equation}

\noindent
Where $\overline{E}(t)$ represents the average energy of electrons or holes at time $t$, $\epsilon_i(t)$ is the energy of the $i$-th adiabatic state $\phi_i(t)$ at time $t$, and $g_i(t)$ is the occupation number of time-dependent wavefunction $\psi_j(t)$ on adiabatic state $\phi_i(t)$ for electrons or holes. $N$ is the total number of states involved in the energy window.

\subsection{Normal-mode coordinates}
The acceleration $\bm{\mathit{a}}_i$ at MD step $t_n$ is approximated as:

\begin{equation}
\bm{\mathit{a}}_i\left( t_n \right) \approx \frac{\mathbf{v}_i\left(t_n\right) - \mathbf{v}_i\left(t_{n-1}\right)}{\Delta t}
\end{equation}

\noindent
where $\mathbf{v}_i\left(t_n\right)$ is the velocity of the $i_{\text{th}}$ atom at MD time $t_n$, and $\Delta t = t_n - t_{n-1}$ is the MD time step. The NOB induced acceleration is the difference between $\bm{\mathit{a}}_i\left( t_n \right)$ and the Verlet-computed acceleration.

\begin{equation}
\bm{\mathit{a}}_i^{\mathrm{NOB}}{\left( t_n \right)} = \bm{\mathit{a}}_i\left( t_n \right) - \frac{\mathbf{F}_i\left(t_n\right)}{m_i}
\end{equation}

\noindent
where $\mathbf{F}_i\left(t_n\right)$ denotes the force on the $i_{\text{th}}$ atom at MD time $t_n$, and $m_i$ is the $i_{\text{th}}$ atom's mass.
\noindent
The NOB event induced velocity increment at step ${t_n}$ will be:
\begin{equation}
\Delta \mathbf{v}_i^{\mathrm{NOB}}(t_n) \approx \bm{\mathit{a}}_i^{\mathrm{NOB}}{\left( t_n \right)} \Delta t
\end{equation}

\noindent
In the following, we want to decompose such velocity increments to the normal mode. The mass-weighted velocity $\mathbf{u}_i(t)$ is: 
\begin{equation}
\mathbf{u}_i(t) = \sqrt{m_i} \Delta \mathbf{v}_i^{\mathrm{NOB}}(t) \ \quad \text{and} \quad \mathbf{u}_{i,\mathbf{q}}(t) = \sum_\mathbf{R} \mathbf{u}_{i,\mathbf{R}}(t) e^{-i \mathbf{q} \cdot \mathbf{R}} \frac{1}{\sqrt{N_p}}
\end{equation}

\noindent
where $\mathbf{q}$ is the phonon wavevector (here taken at the $\Gamma$-point), $\mathbf{R}$ is a unitcell lattice site in the supercell, and $\mathbf{u}_{i,\mathbf{q}}(t)$ is the real-space Fourier transform of $\mathbf{u}_{i,\mathbf{R}}(t)$. $N_p$ is the number of unit cells inside the supercell. In our case, $N_p$ = 1.

\begin{equation}
P_\nu (t) = \sum_{i} \left| \mathbf{u}_{i,\mathbf{q}}(t) \cdot \mathbf{e}^*_{i,\nu}(\mathbf{q}) \right|^2
\end{equation}

\noindent
where $\mathbf{e}^*_{i,\nu}(\mathbf{q})$ is the normalized eigenvector of the dynamical matrix for phonon mode $\nu$ on the $i_{\text{th}}$ atom. Here, $P_\nu (t)$ is proportional to the kinetic energy decomposed to normal mode $\nu$.

\begin{figure}[H]
\centering
\includegraphics[width=0.8\textwidth]{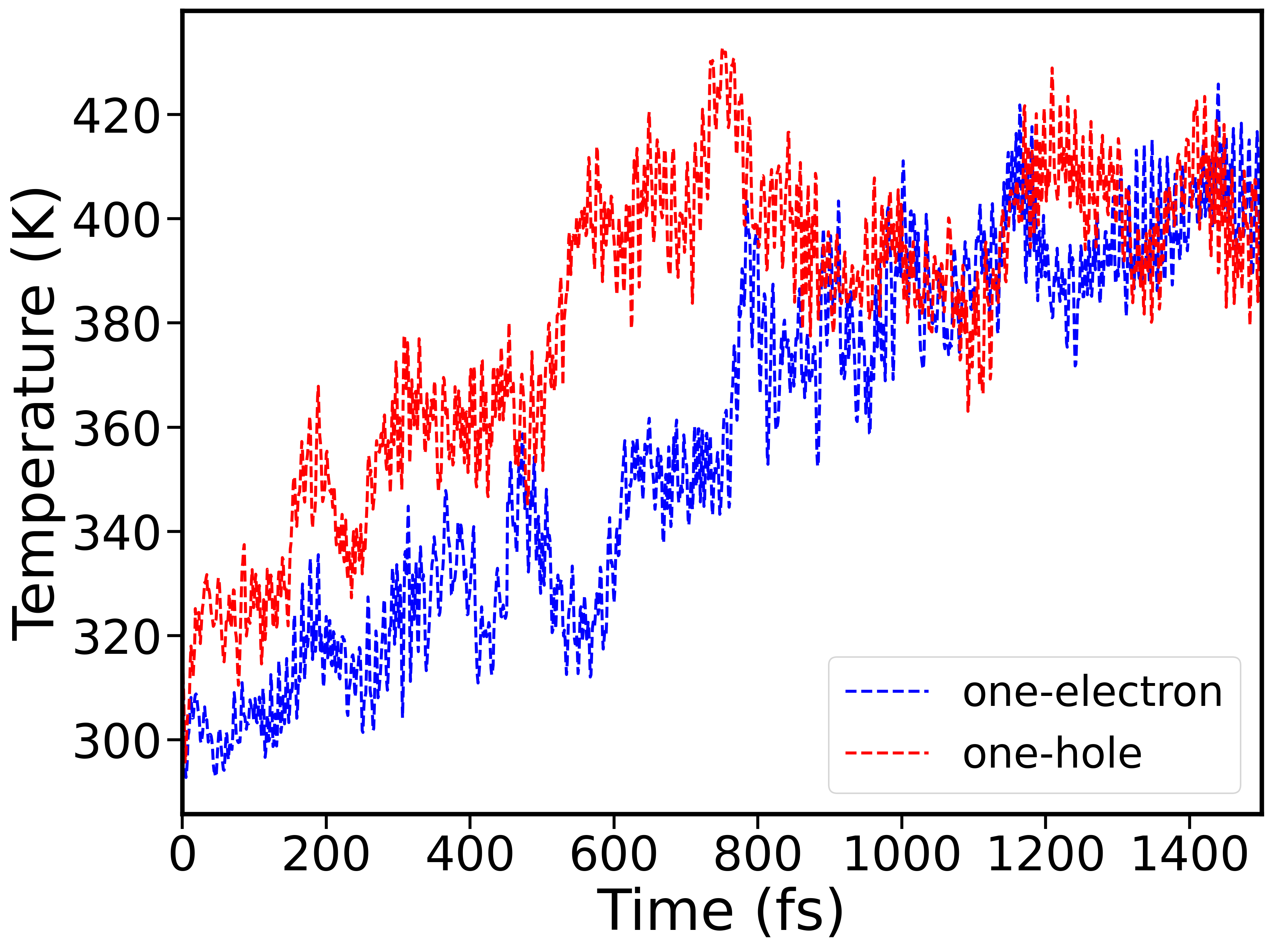}
\renewcommand{\thefigure}{S\arabic{figure}}
\caption{Cooling process of a single electron or hole and the corresponding temperature change in the system.}
\label{fig:temp-cooling}
\end{figure}

\begin{figure}[H]
\centering
\includegraphics[width=1\textwidth]{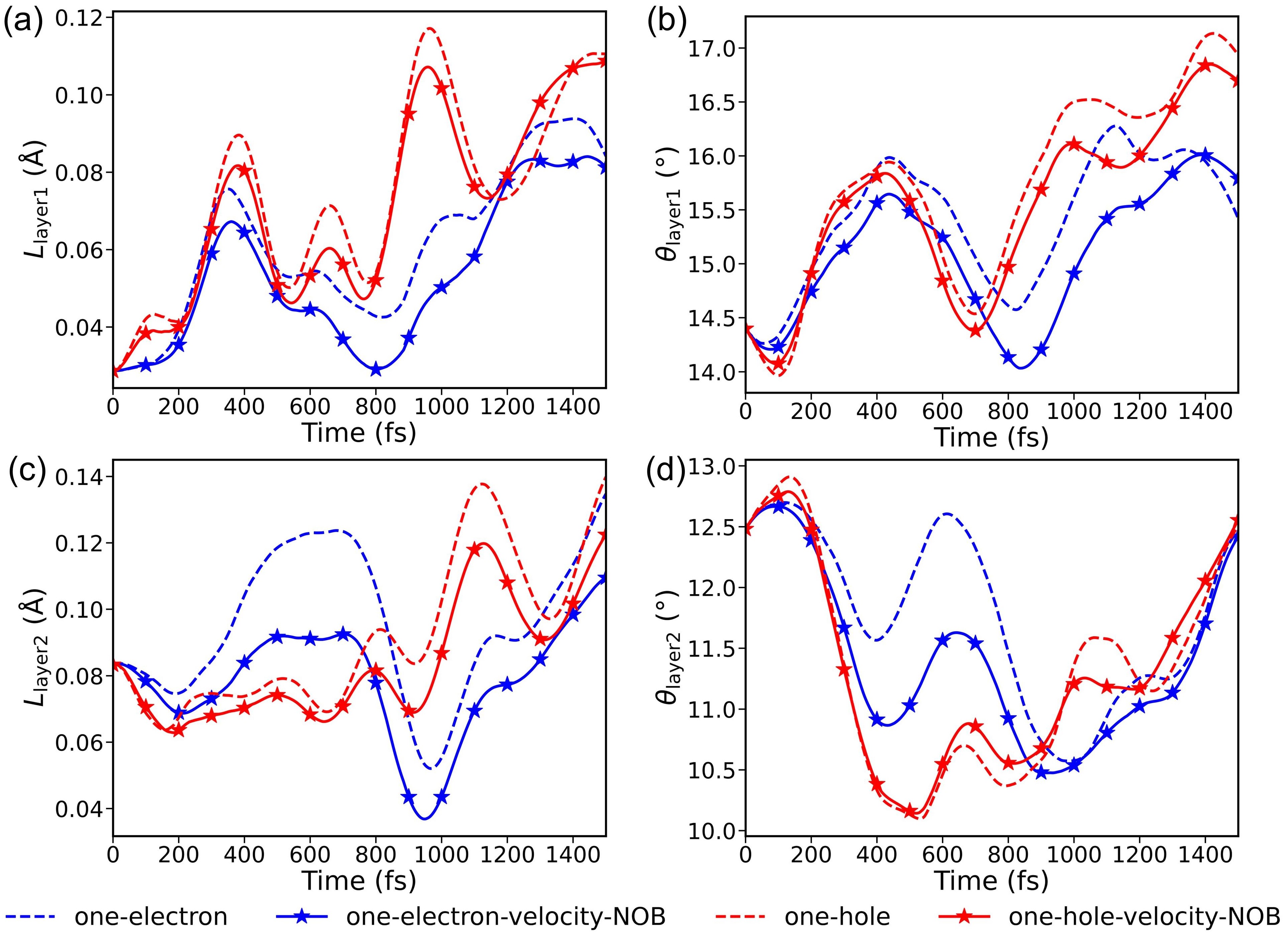}
\renewcommand{\thefigure}{S\arabic{figure}}
\caption{$L$ (Å) and $\theta$ (°) for different layers. “-velocity-NOB” refers to a special BOMD simulation in which additional velocities generated at the moment of a NOB event from the TDDFT-NOB trajectory—are added to the corresponding atoms at the same time step in the BOMD trajectory.}
\label{fig:velocity_vector}
\end{figure}

\begin{figure}[H]
\centering
\includegraphics[width=1\textwidth]{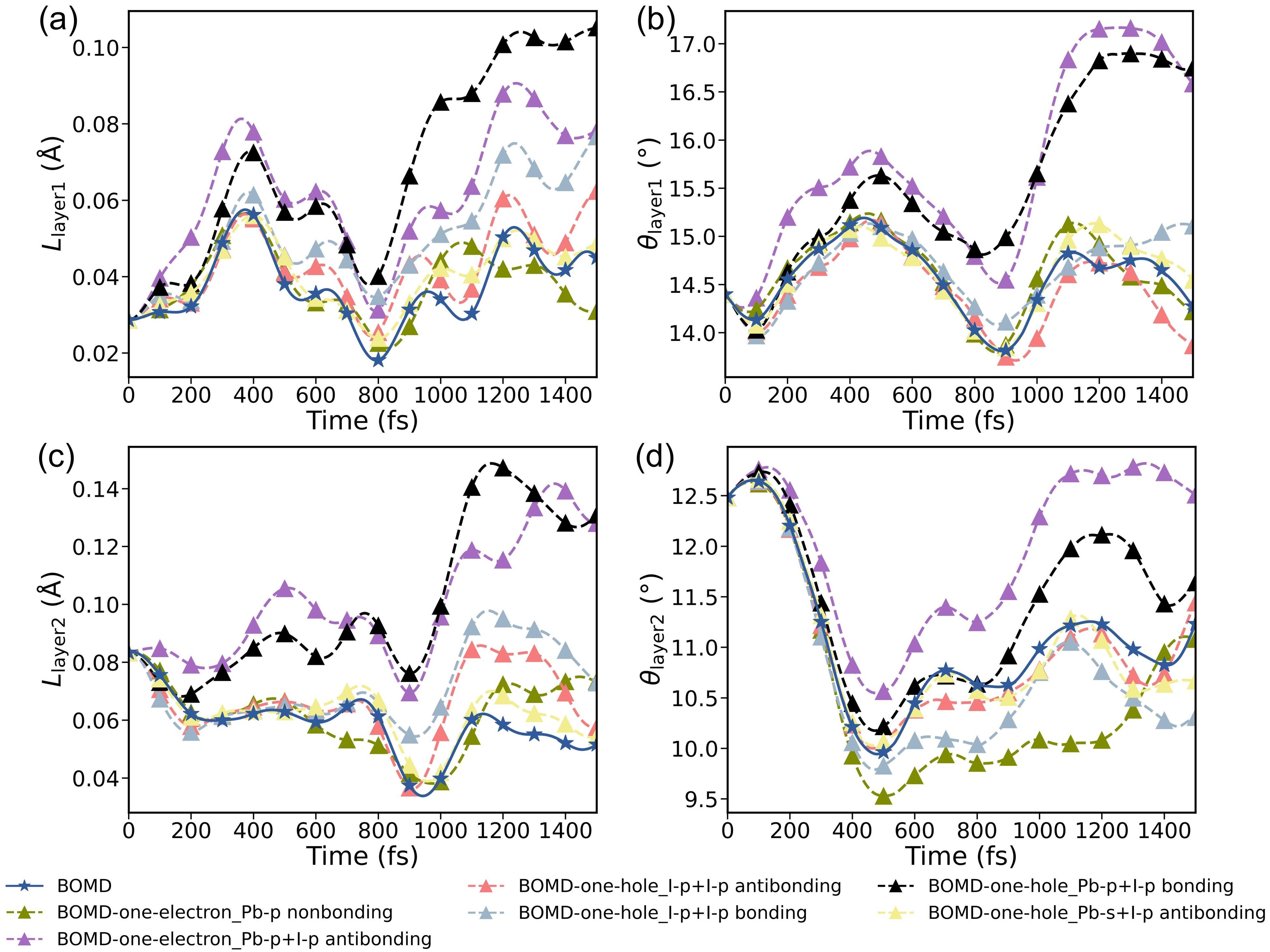}
\renewcommand{\thefigure}{S\arabic{figure}}
\caption{$L$ (\r{A}) and $\theta$ ($^\circ$) for different layers. Lines with stars represent ground-state BOMD, while colored triangles denote states associated with distinct orbital character regions shown in Fig.1 Each region is assigned one electron or one hole in total.}
\label{fig:constarined_vector}
\end{figure}

\begin{figure}[H]
\centering
\includegraphics[width=1\textwidth]{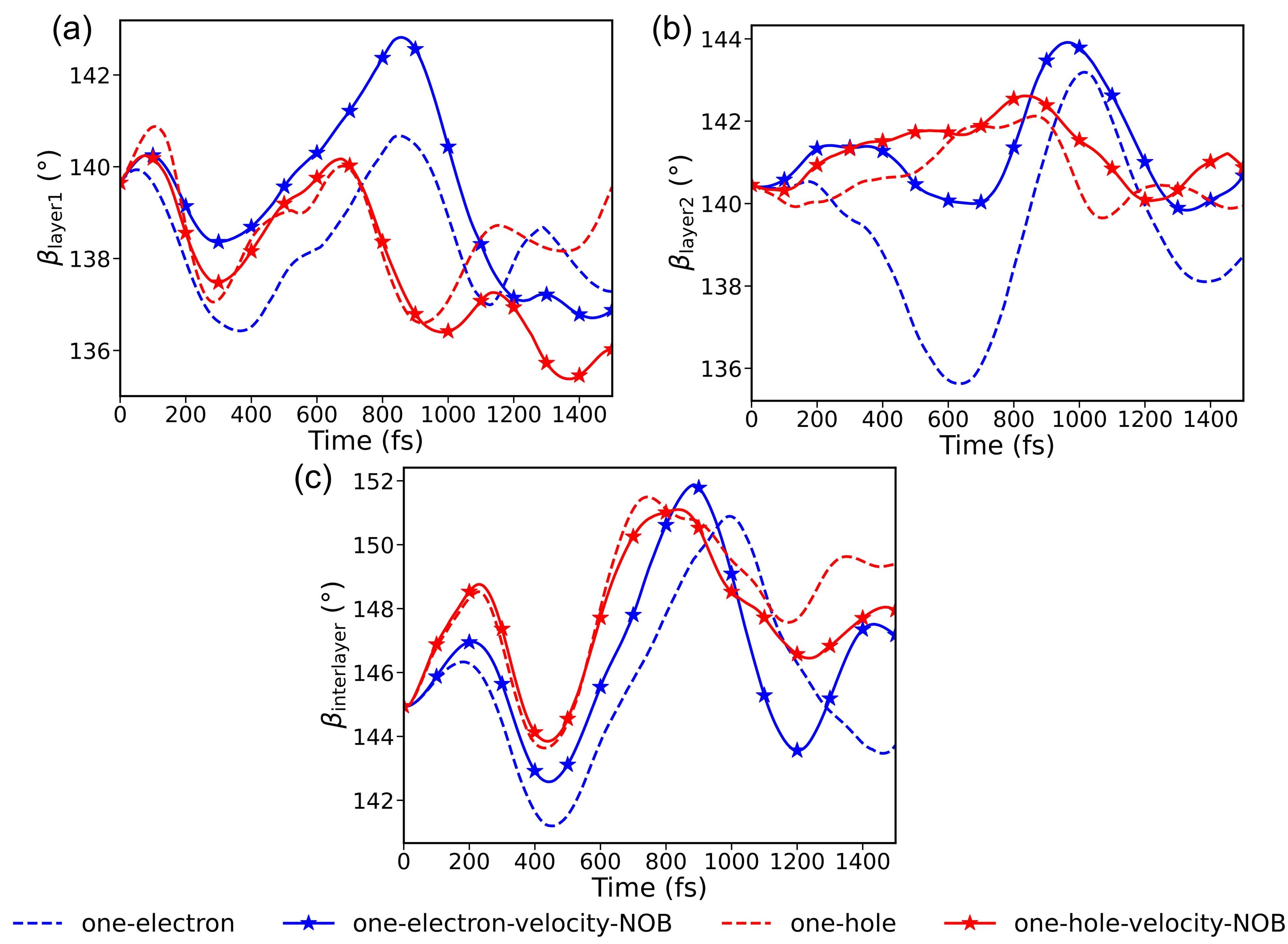}
\renewcommand{\thefigure}{S\arabic{figure}}
\caption{$\beta$ ($^\circ$) for different layers. “-velocity-NOB” refers to a special BOMD simulation in which additional velocities generated at the moment of a NOB event from the TDDFT-NOB trajectory—are added to the corresponding atoms at the same time step in the BOMD trajectory.}
\label{fig:velocity_angle}
\end{figure}

\begin{figure}[H]
\centering
\includegraphics[width=1\textwidth]{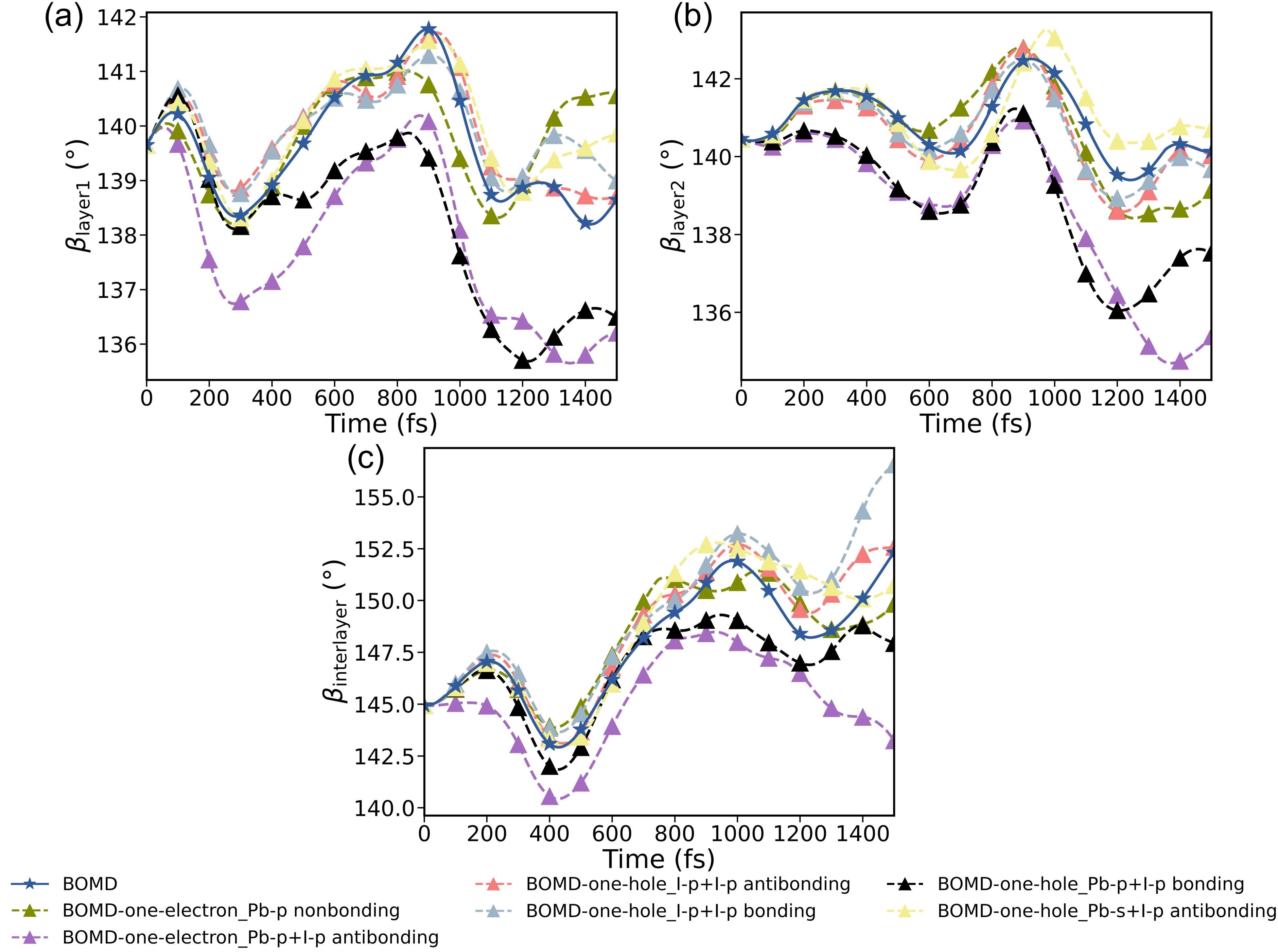}
\renewcommand{\thefigure}{S\arabic{figure}}
\caption{$\beta$ ($^\circ$) for different layers. Lines with stars represent ground-state BOMD, while colored triangles denote states associated with distinct orbital character regions shown in Fig.1 Each region is assigned one electron or one hole in total.}
\label{fig:constarined_angle}
\end{figure}
\renewcommand{\thetable}{S\arabic{table}}

\section{The Pb$_2$I$_2$ quadrilateral}

\begin{figure}[H]
\centering
\includegraphics[width=0.8\textwidth]{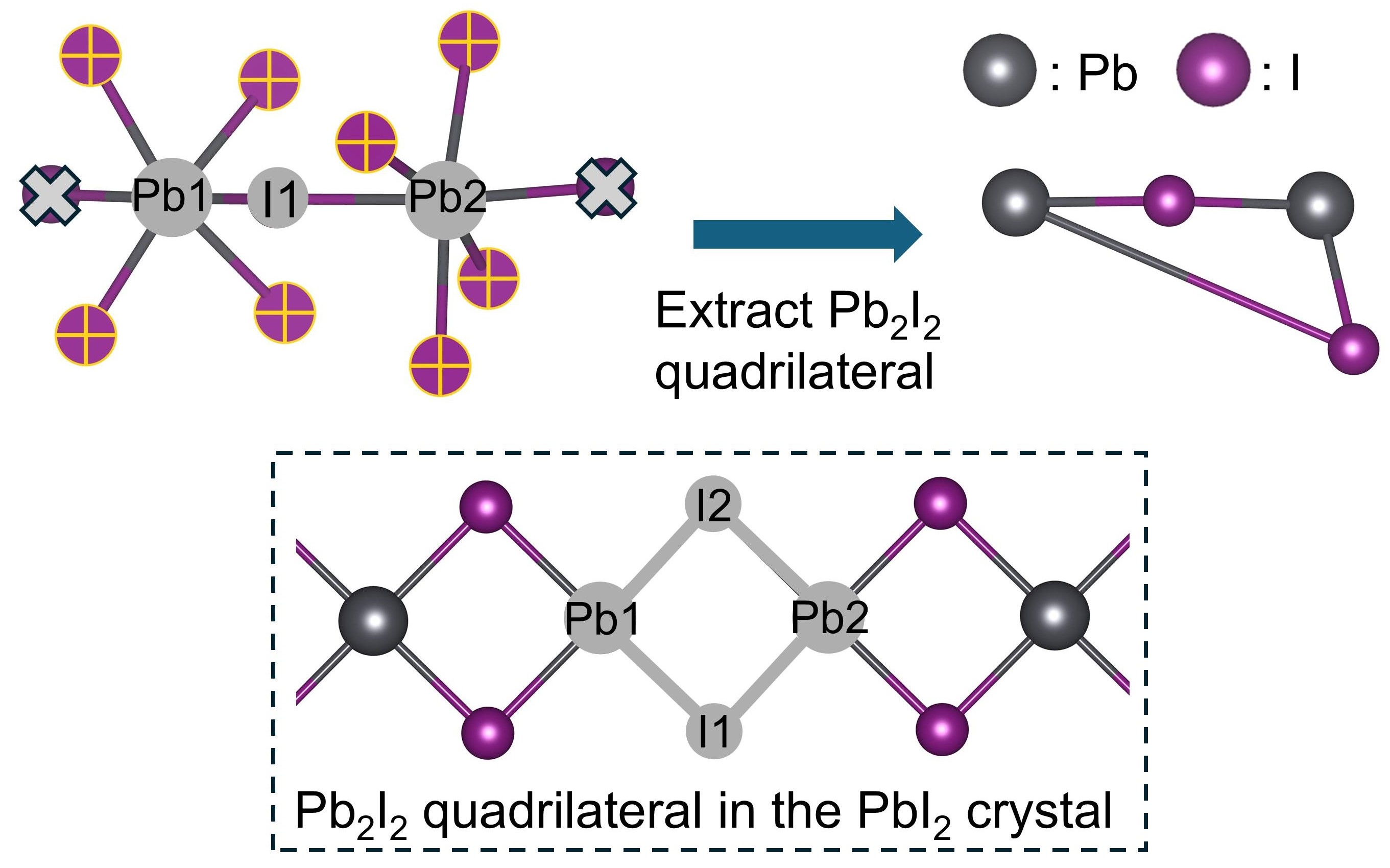}
\renewcommand{\thefigure}{S\arabic{figure}}
\caption{For each Pb--I--Pb angle, I atoms with a distance less than 5.5~\AA\ to the Pb--I--Pb angle (highlighted I atoms, with a cross symbol indicating I atoms further than 5.5~\AA\ are not considered) are selected to form the Pb$_2$I$_2$ quadrilateral. The Pb$_2$I$_2$ quadrilateral in perfect PbI$_2$ is also shown.}
\label{fig:Pb2I2-show}
\end{figure}

To identify Pb$_2$I$_2$ quadrilaterals, we begin by examining each Pb-I-Pb angle, which involves two Pb atoms and a bridging I atom. For each Pb-I-Pb angle, all coordinating I atoms to these two Pb atoms are searched and the I atom with a shortest distance (such distance must be less than 5.5~\AA) to the bridging I atom is selected (Fig.\ref{fig:Pb2I2-show}). The selected I atom plus the three Pb-I-Pb atoms compose the Pb$_2$I$_2$ quadrilateral. Next, we remove duplicate quadrilaterals and calculate the distances between the six specified distances in each quadrilateral: four Pb-I distances, Pb-Pb distance along the diagonal divided by $\sqrt{2}$, and I-I distance along the diagonal divided by $\sqrt{2}$. The absolute difference of each distance from the average are calculated and the mean of six differences is taken as the variation of the quadrilateral, denoted as $\delta$. For perfect PbI$_2$, $\delta = 0$ \AA. For equilibrium MAPbI$_3$ lattice, the calculated $\delta$ is around 0.6 \AA.  A smaller value of $\delta$ indicates that the Pb$_2$I$_2$ quadrilateral in MAPbI$_3$ is closer to that in PbI$_2$ lattice.

\begin{figure}[H]
\centering
\includegraphics[width=0.8\textwidth]{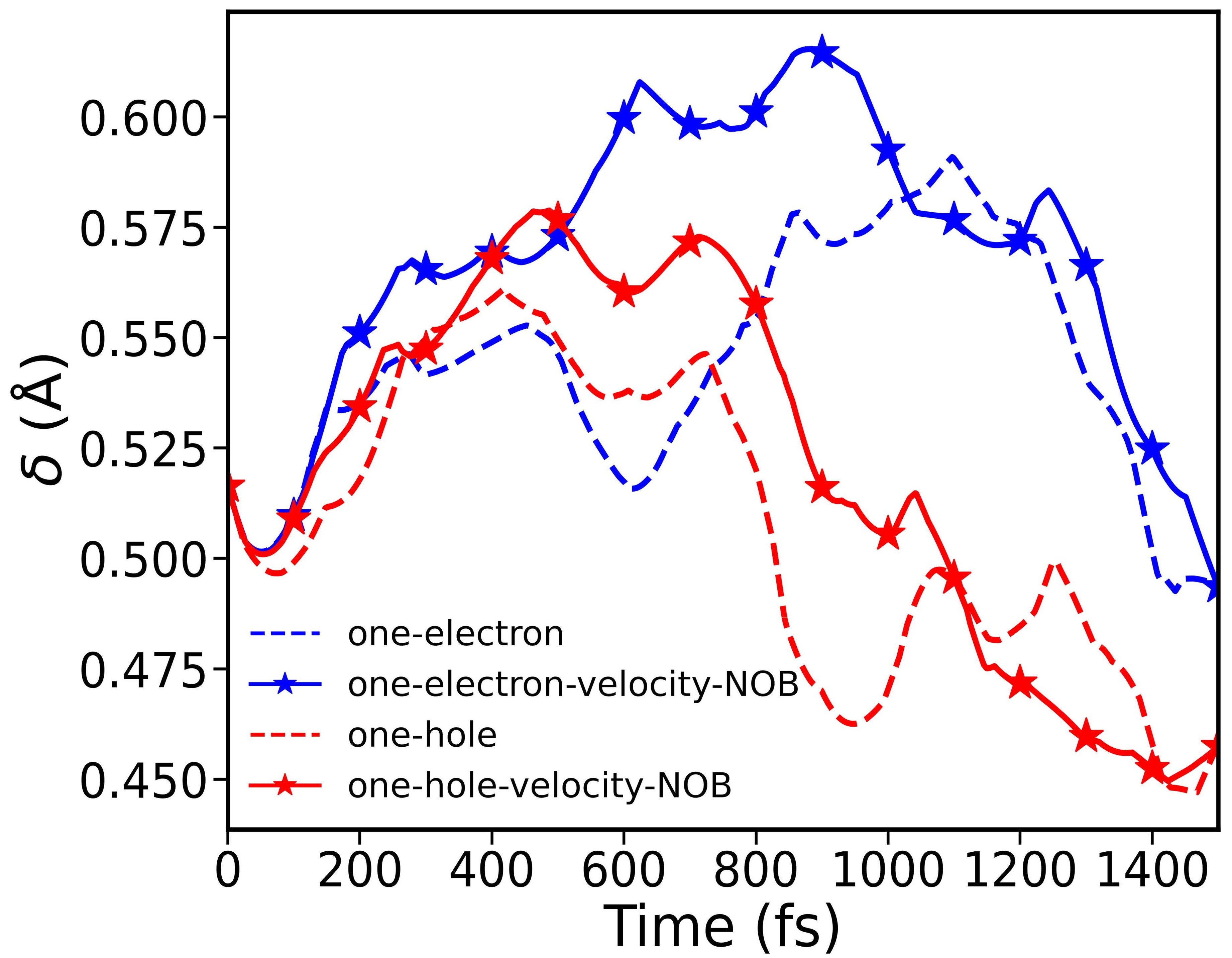}
\renewcommand{\thefigure}{S\arabic{figure}}
\caption{The evolution of $\delta$ (\AA). “-velocity-NOB” refers to a special BOMD simulation in which additional velocities generated at the moment of a NOB event from the TDDFT-NOB trajectory—are added to the corresponding atoms at the same time step in the BOMD trajectory.}
\label{fig:velocity_min}
\end{figure}

\begin{figure}[H]
\centering
\includegraphics[width=1\textwidth]{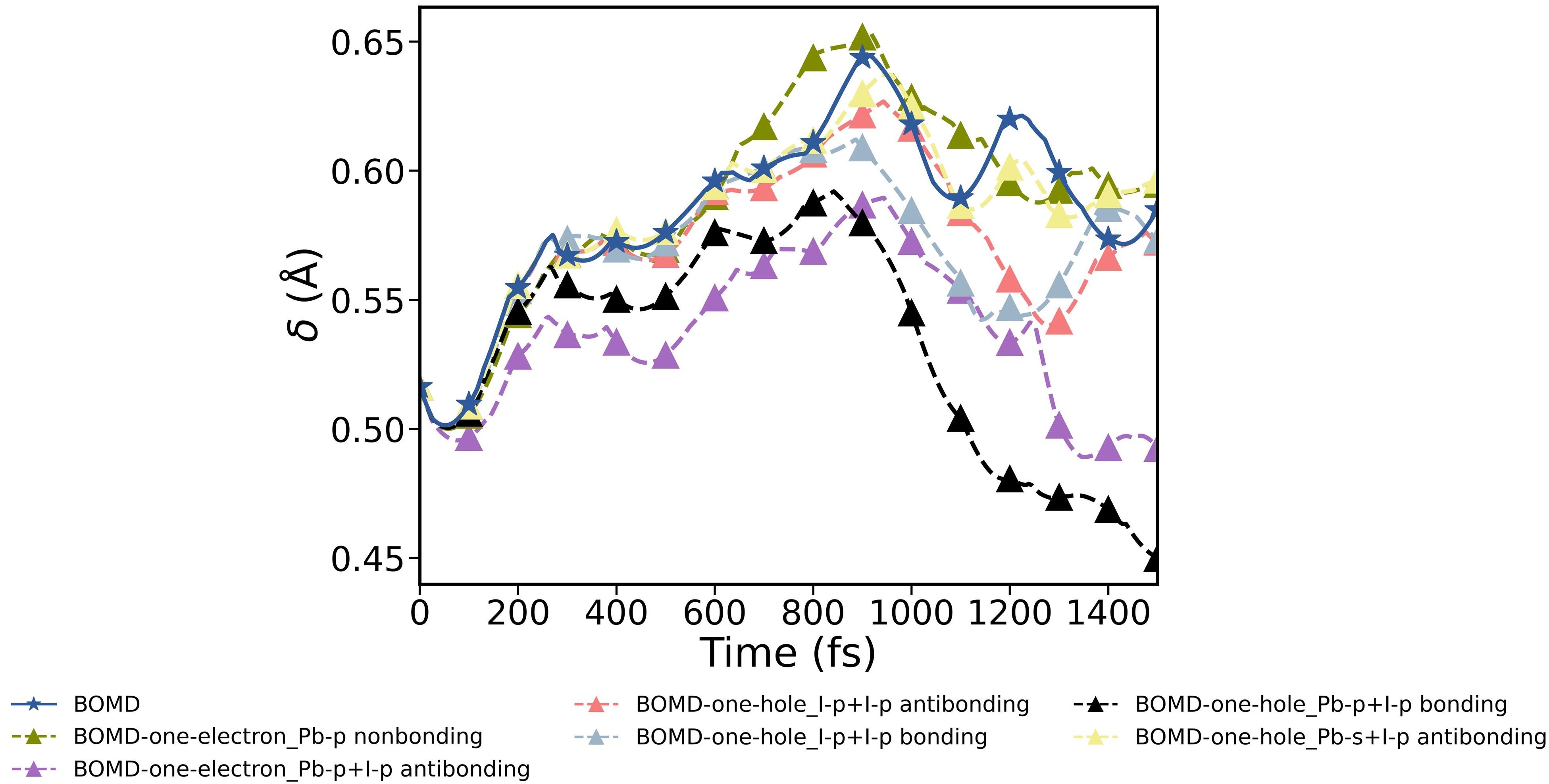}
\renewcommand{\thefigure}{S\arabic{figure}}
\caption{The evolution of $\delta$ (\AA). Lines with stars represent ground-state BOMD, while colored triangles denote states associated with distinct orbital character regions shown in Fig.1 Each region is assigned one electron or one hole in total.}
\label{fig:constarined_min}
\end{figure}
\renewcommand{\thetable}{S\arabic{table}}

\begin{table}[H]
\centering
\renewcommand{\arraystretch}{1.3}
\begin{tabular}{lccc}
\hline
 & C--N$_\mathrm{max}$ (Å) & C--H$_\mathrm{max}$ (Å) & N--H$_\mathrm{max}$ (Å) \\
\hline
1 electron & 1.600 & 1.174 & 1.102 \\
1 hole     & 1.562 & 1.166 & 1.119 \\
\hline
\end{tabular}
\caption{Maximum C--N, C--H, and N--H bond lengths in MA$^+$ under one-electron or one-hole excitation.}
\label{tab:max_bond_lengths}
\end{table}

\begin{figure}[H]
\centering
\includegraphics[width=0.8\textwidth]{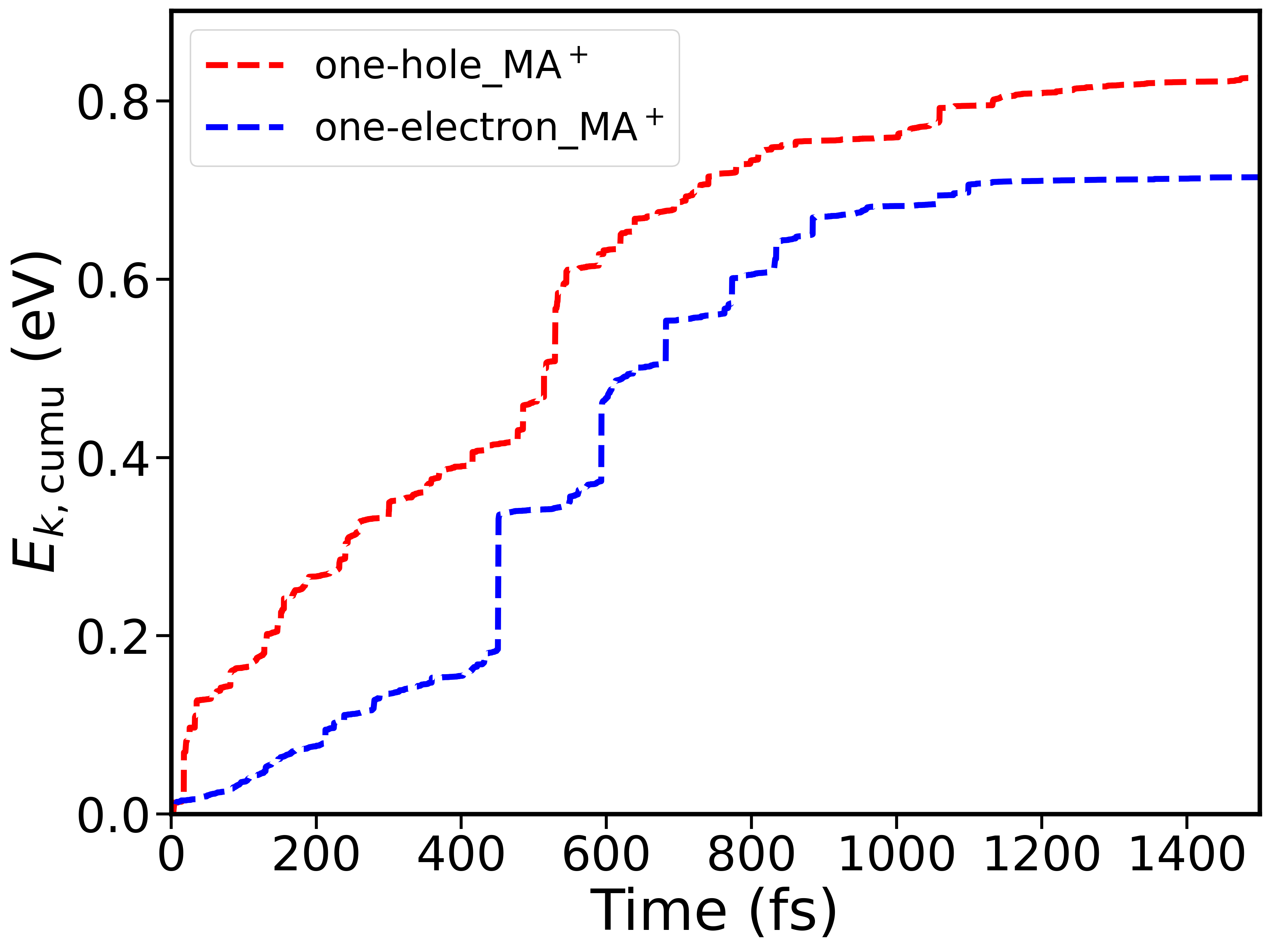}
\renewcommand{\thefigure}{S\arabic{figure}}
\caption{Cumulative kinetic energy ($E_{k,cumu}$) (Equ.S10) transferred to MA$^+$ by NOB events.}
\label{fig:MA-analyze}
\end{figure}

\begin{equation}
E_{k,cumu}(t) = \sum_{t_0}^{t} \sum_{i \in \mathrm{MA}^+} \frac{1}{2} m_i \left| \Delta \mathbf{v}_i^{\mathrm{NOB}}(t) \right|^2
\end{equation}

\( E_{k,cumu}(t) \) is the cumulative kinetic energy which sums the kinetic energy from the first NOB event at \( t_0 \) to the current time step \( t \). The second summation over \( i \) runs over all atoms that belong to the MA\(^+\) group. The term \( \frac{1}{2} m_i \left|\Delta \mathbf{v}_i^{\mathrm{NOB}}(t)\right|^2 \) represents the kinetic energy of each atom, where \( \Delta \mathbf{v}_i^{\mathrm{NOB}}(t) \) is the velocity increment at time \( t \) induced by the NOB event, and it is calculated using Equation S7.

\begin{figure}[H]
\centering
\includegraphics[width=1\textwidth]{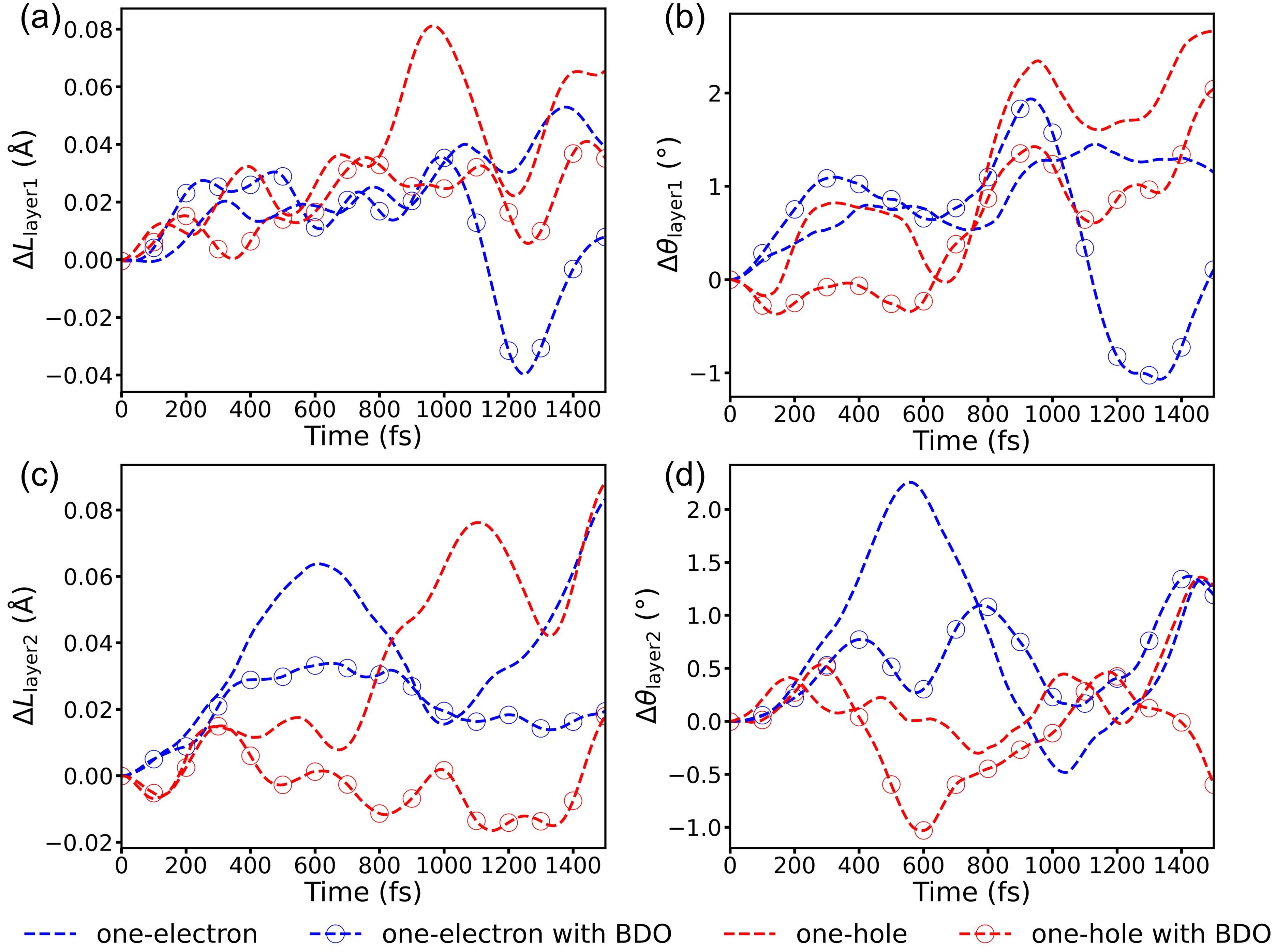}
\renewcommand{\thefigure}{S\arabic{figure}}
\caption{Time evolution of $\Delta L$ and $\Delta \theta$ for layer~1 and layer~2. The corresponding changes in MAPbI$_3$ with BDO are represented by lines with circles.}
\label{fig:BDO-vector}
\end{figure}

\begin{figure}[H]
\centering
\includegraphics[width=0.8\textwidth]{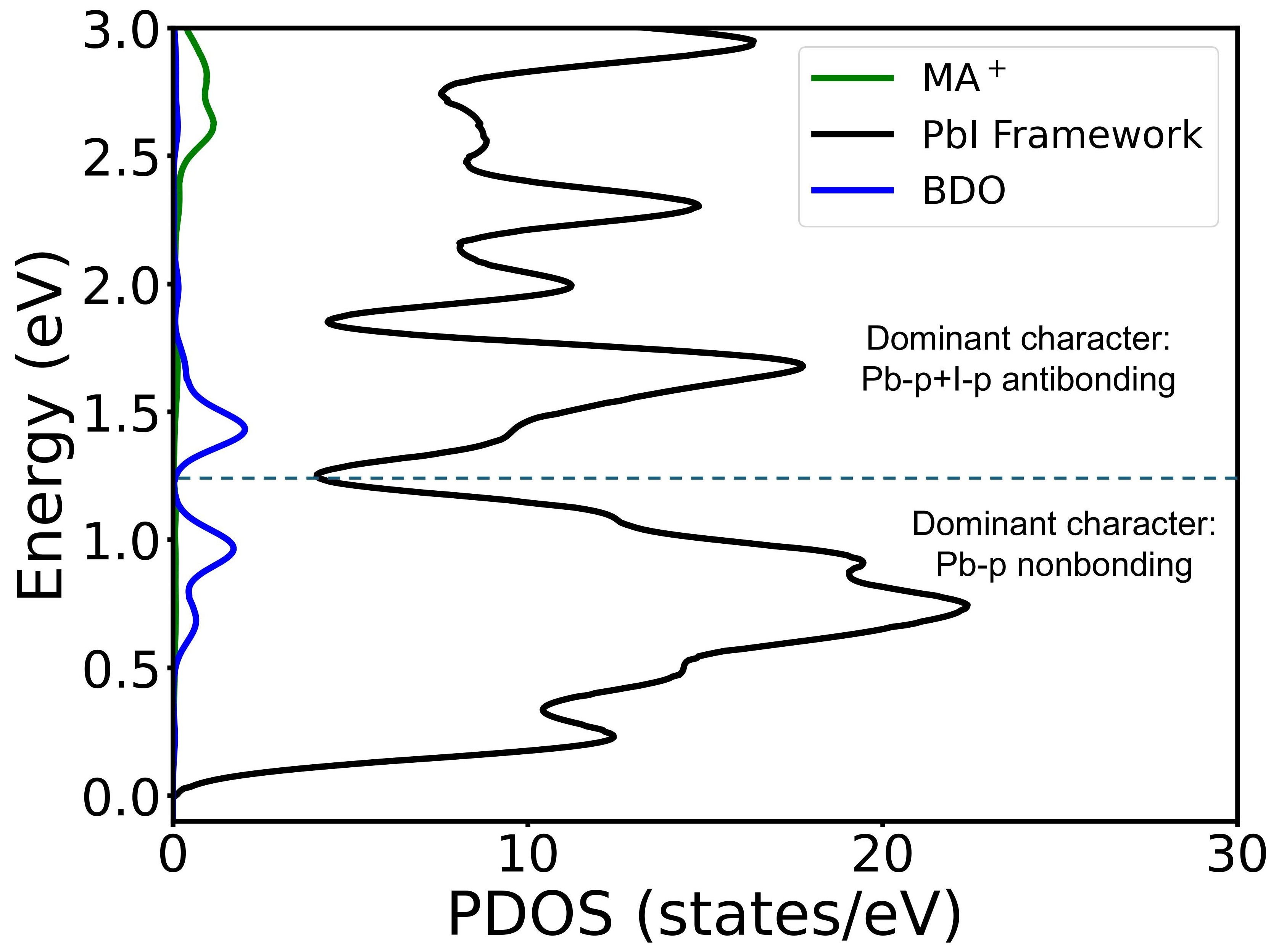}
\renewcommand{\thefigure}{S\arabic{figure}}
\caption{Density of states (DOS) of different types of atoms in MAPbI$_3$ with adsorbed BDO.}
\label{fig:DOS-BDO}
\end{figure}

\begin{figure}[H]
\centering
\includegraphics[width=1\textwidth]{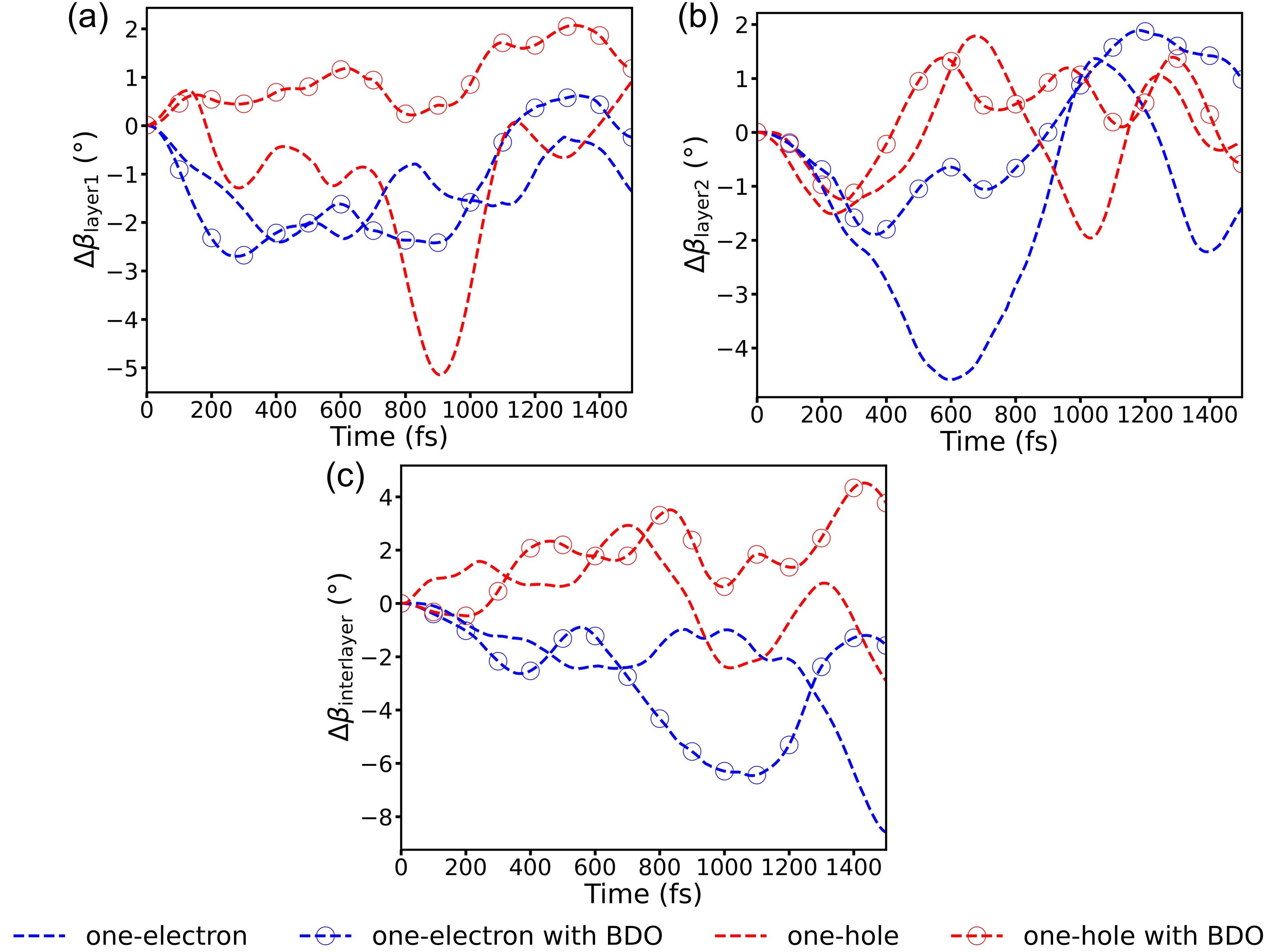}
\renewcommand{\thefigure}{S\arabic{figure}}
\caption{Time evolution of $\Delta \beta$ for layer~1, layer~2, and the interlayer (Pb--I--Pb angle connecting five- and six-coordinated polyhedra). The corresponding changes in MAPbI$_3$ with BDO are represented by lines with circles.}
\label{fig:BDO-angle}
\end{figure}

\begin{figure}[H]
\centering
\includegraphics[width=0.8\textwidth]{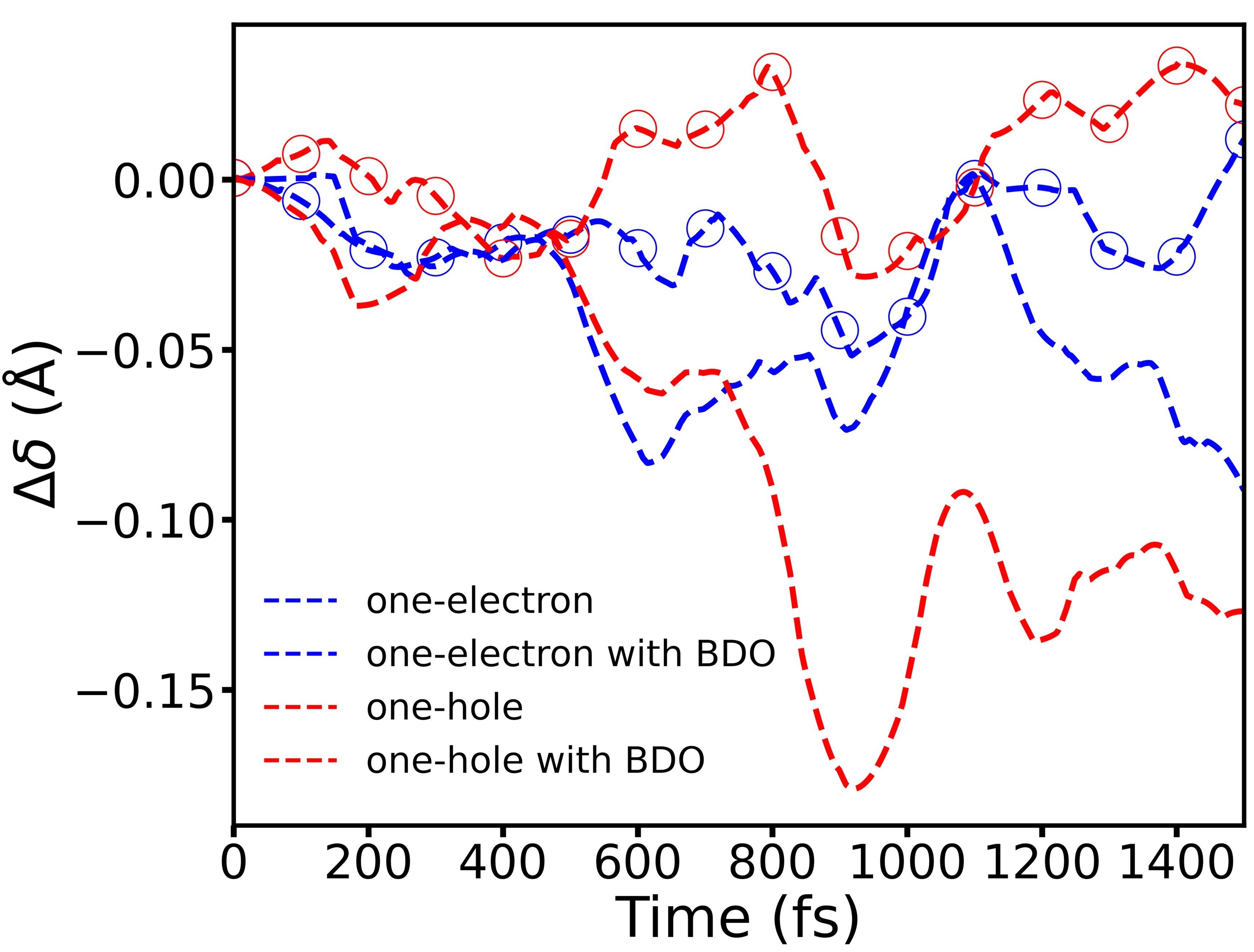}
\renewcommand{\thefigure}{S\arabic{figure}}
\caption{The evolution of $\Delta \delta$. The corresponding changes in MAPbI$_3$ with BDO are represented by lines with circles.}
\label{fig:BDO-min}
\end{figure}

\begin{figure}[H]
\centering
\includegraphics[width=0.8\textwidth]{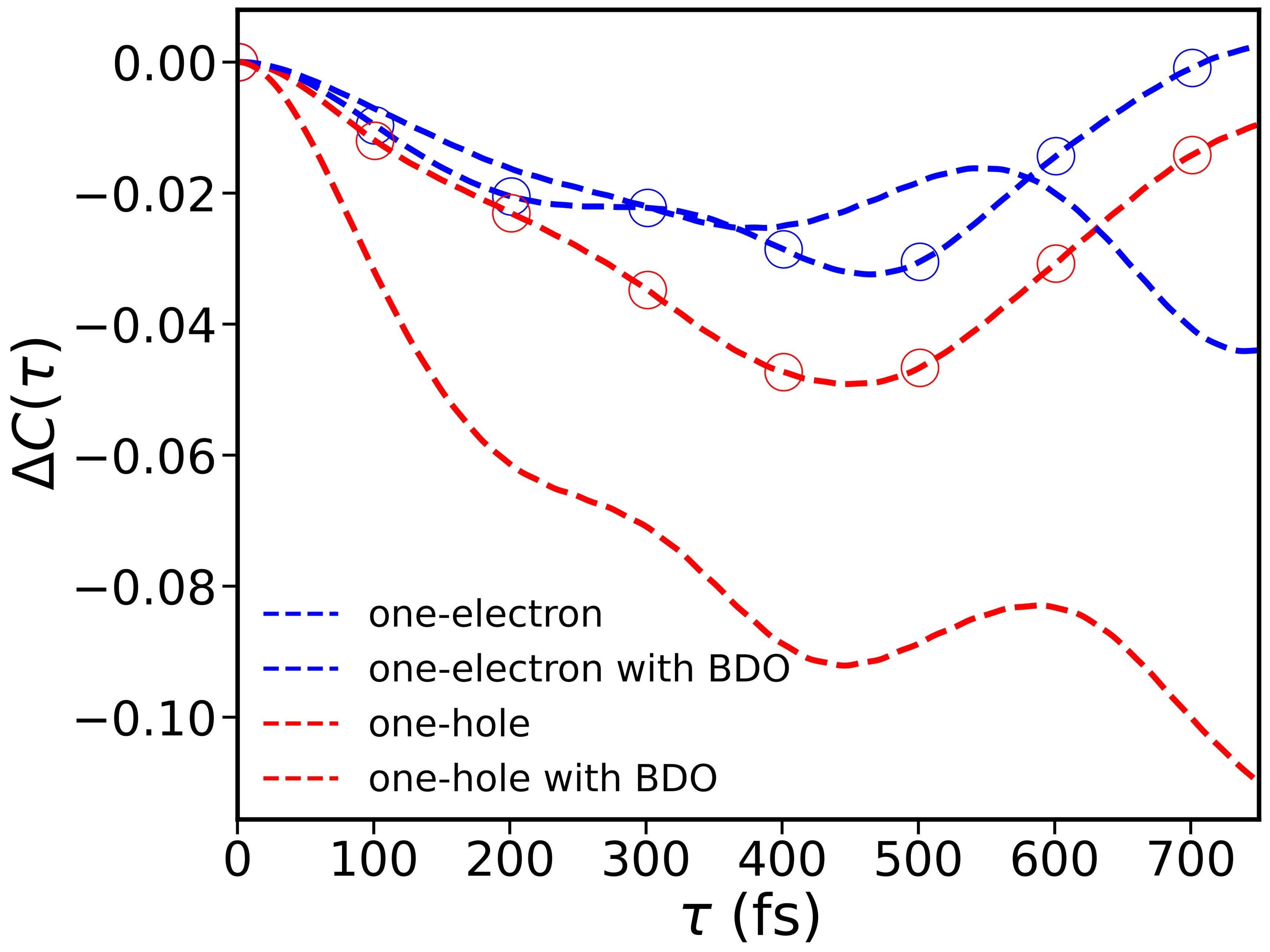}
\renewcommand{\thefigure}{S\arabic{figure}}
\caption{The autocorrelation function of MA$^+$ orientation as a function of time. The corresponding changes in MAPbI$_3$ with BDO are represented by lines with circles.}
\label{fig:MA_BDO}
\end{figure}

\begin{figure}[H]
\centering
\includegraphics[width=1\textwidth]{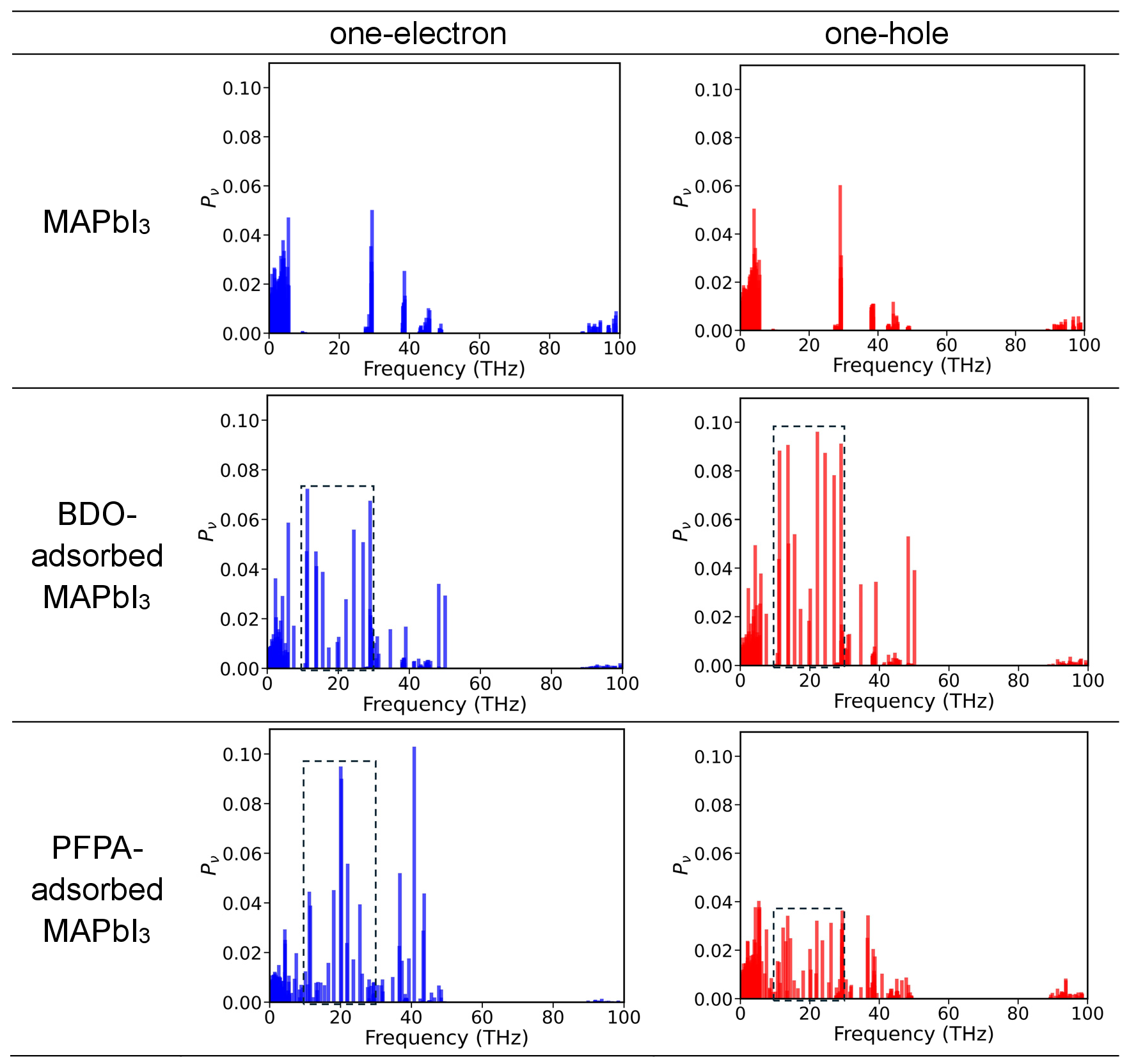}
\renewcommand{\thefigure}{S\arabic{figure}}
\caption{Accumulate $P_\nu$ from each NOB event for each phonon mode. The dashed box indicates phonon modes that are not present in MAPbI$_3$ but are associated with the passivator. The method to compute $P_\nu$ is given in Equ.S5 to S9.}
\label{fig:PFPA-BDO-projection}
\end{figure}

\begin{figure}[H]
\centering
\includegraphics[width=1\textwidth]{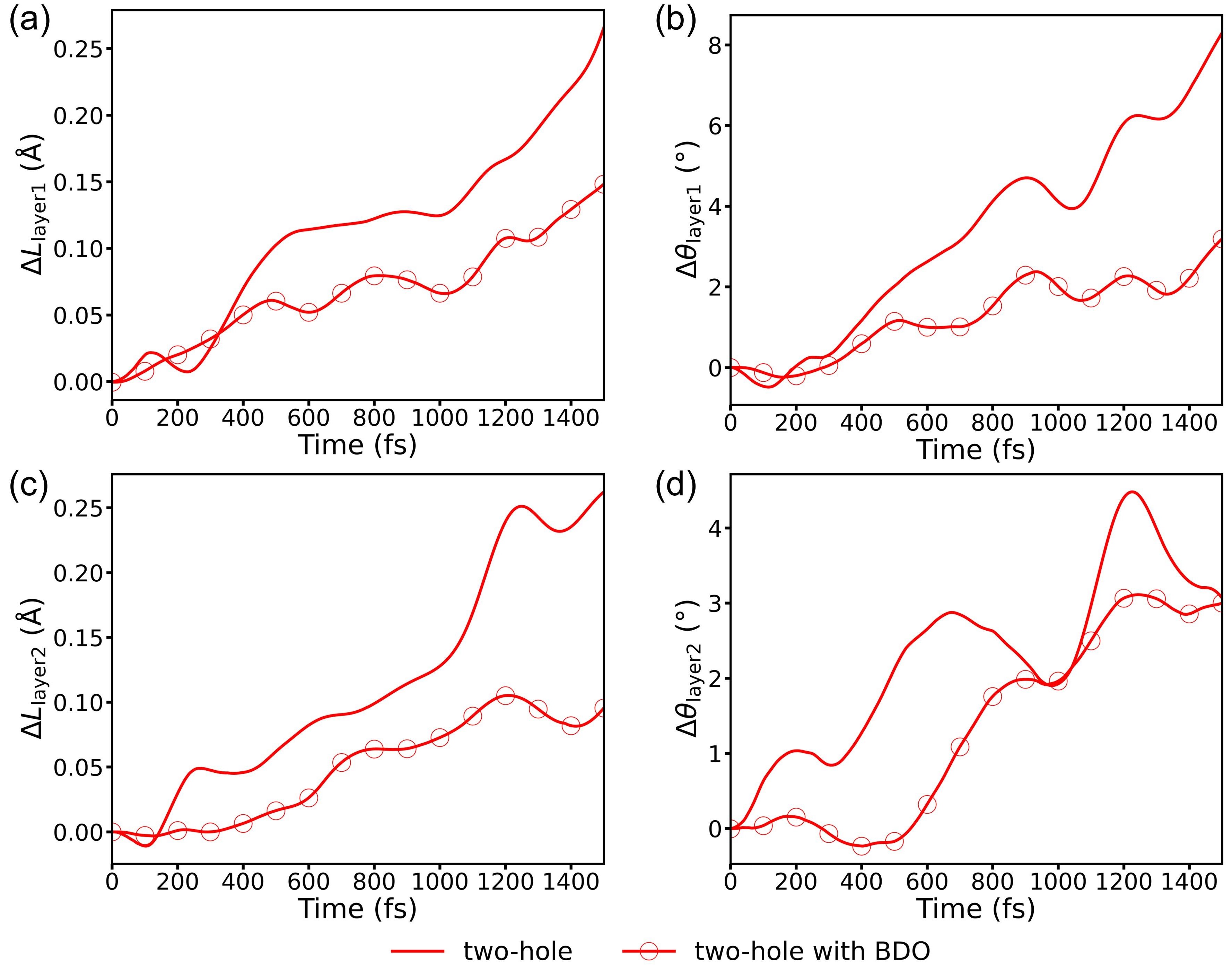}
\renewcommand{\thefigure}{S\arabic{figure}}
\caption{Time evolution of $\Delta L$ and $\Delta \theta$ for layer~1 and layer~2. The corresponding changes in MAPbI$_3$ with BDO are represented by lines with circles.}
\label{fig:BDO-vector-2hole}
\end{figure}

\begin{figure}[H]
\centering
\includegraphics[width=1\textwidth]{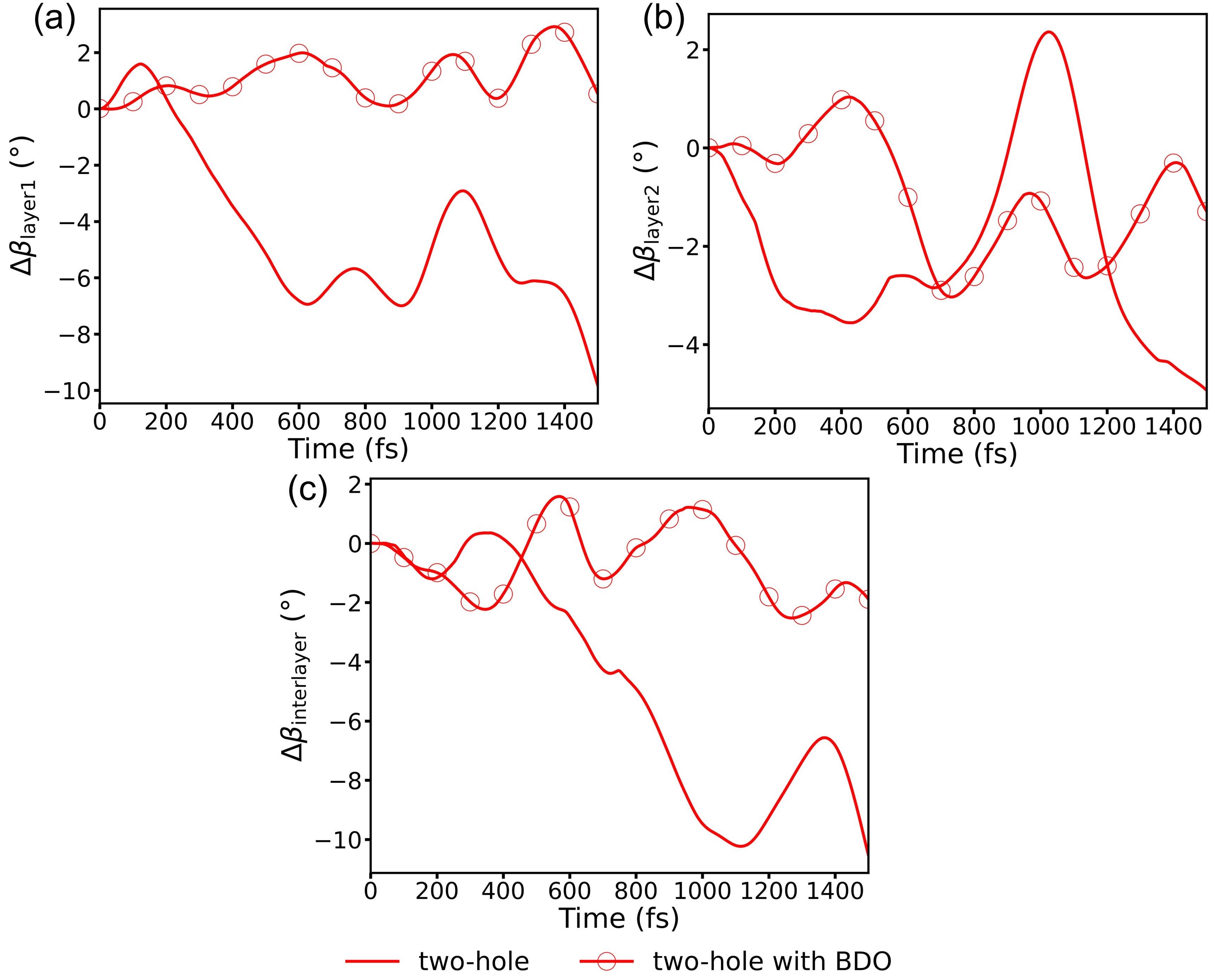}
\renewcommand{\thefigure}{S\arabic{figure}}
\caption{Time evolution of $\Delta \beta$ for layer~1, layer~2, and the interlayer (Pb--I--Pb angle connecting five- and six-coordinated polyhedra). The corresponding changes in MAPbI$_3$ with BDO are represented by lines with circles.}
\label{fig:BDO-angle-2hole}
\end{figure}

\begin{figure}[H]
\centering
\includegraphics[width=0.8\textwidth]{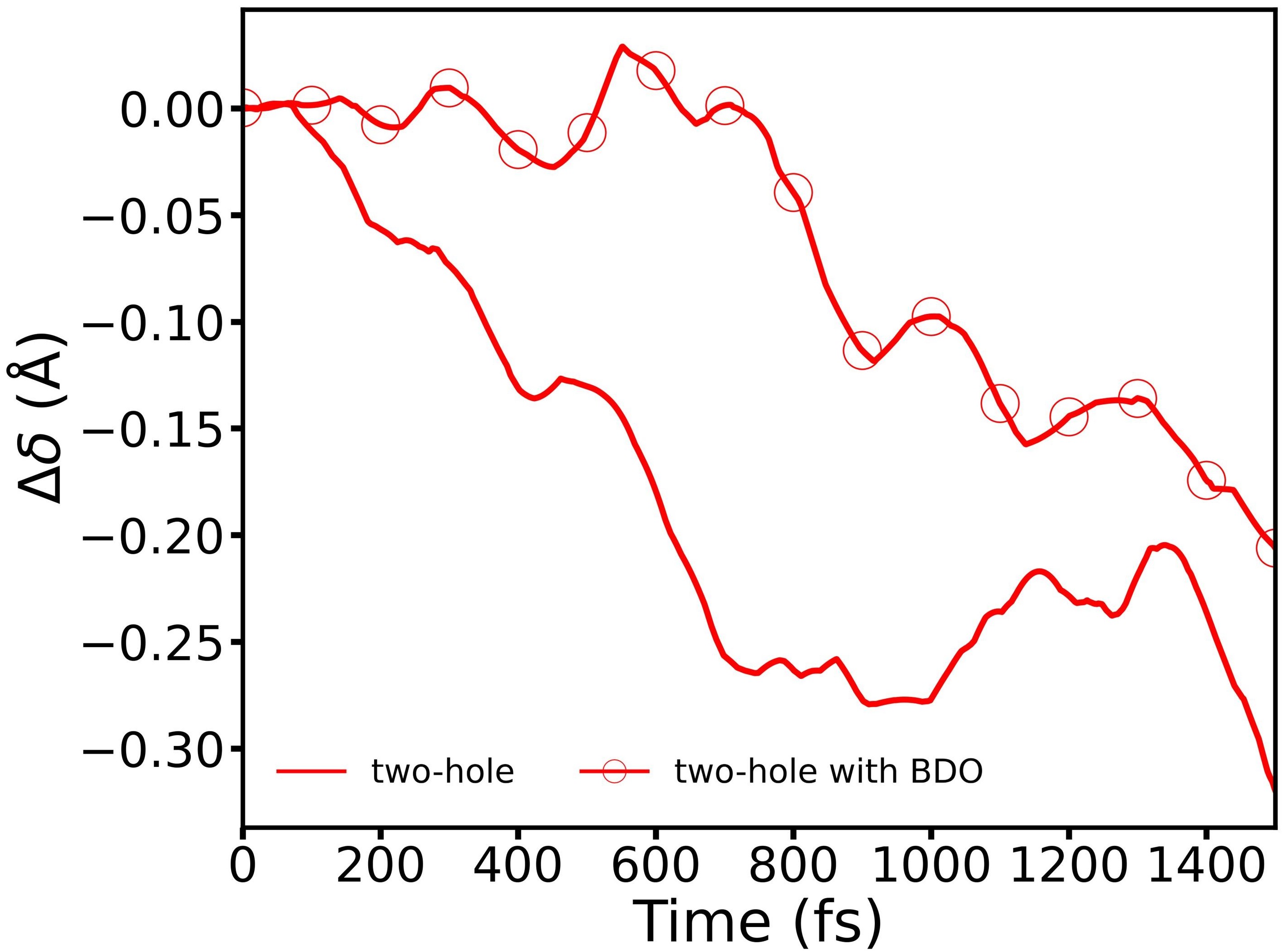}
\renewcommand{\thefigure}{S\arabic{figure}}
\caption{The evolution of $\Delta \delta$. The corresponding changes in MAPbI$_3$ with BDO are represented by lines with circles.}
\label{fig:BDO-min-2hole}
\end{figure}

\begin{figure}[H]
\centering
\includegraphics[width=0.8\textwidth]{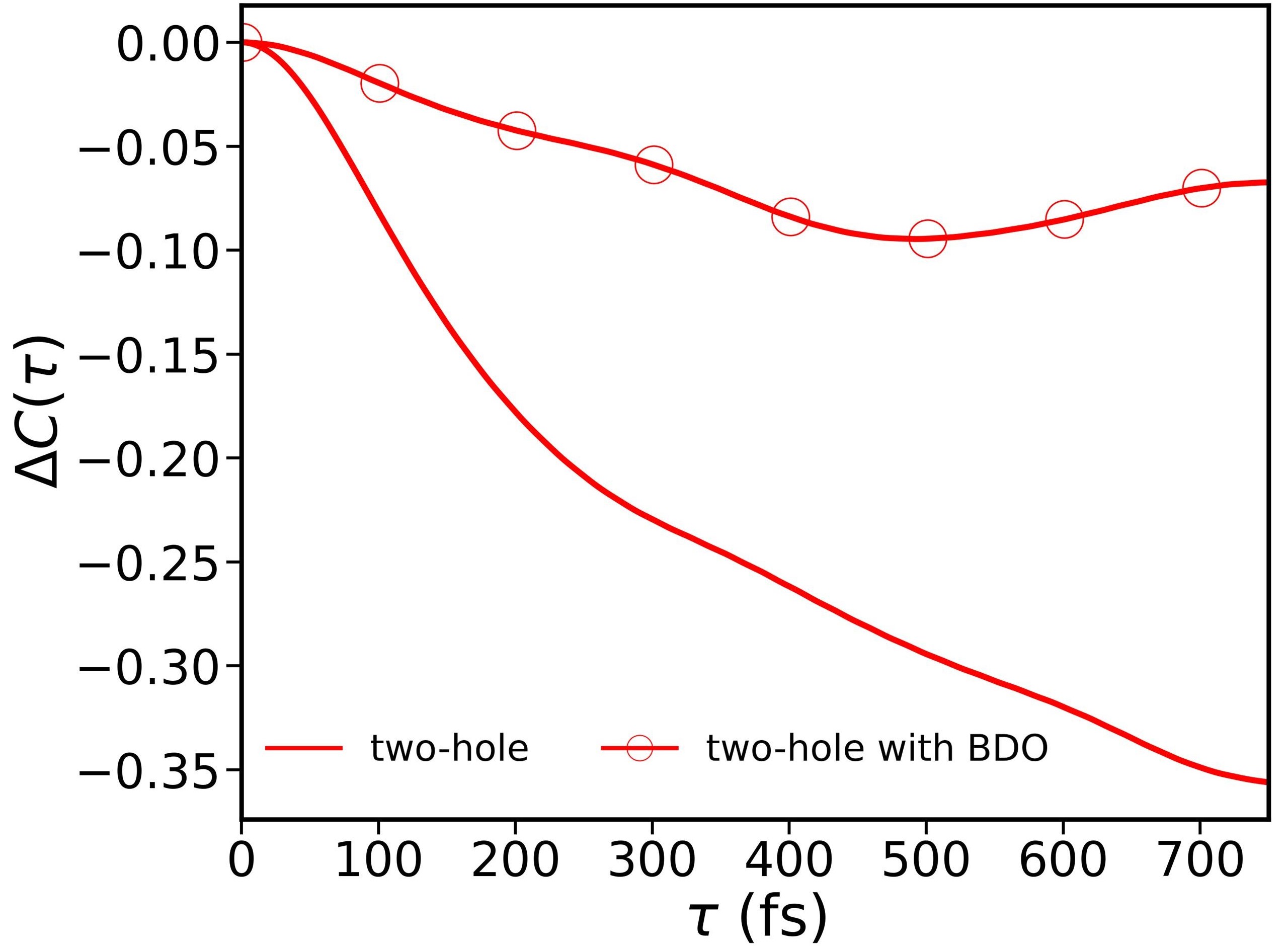}
\renewcommand{\thefigure}{S\arabic{figure}}
\caption{The evolution of the autocorrelation function of MA$^+$ at different time intervals $\tau$. The corresponding changes in MAPbI$_3$ with BDO are represented by lines with circles.}
\label{fig:BDO-correlation}
\end{figure}

\begin{figure}[H]
\centering
\includegraphics[width=1\textwidth]{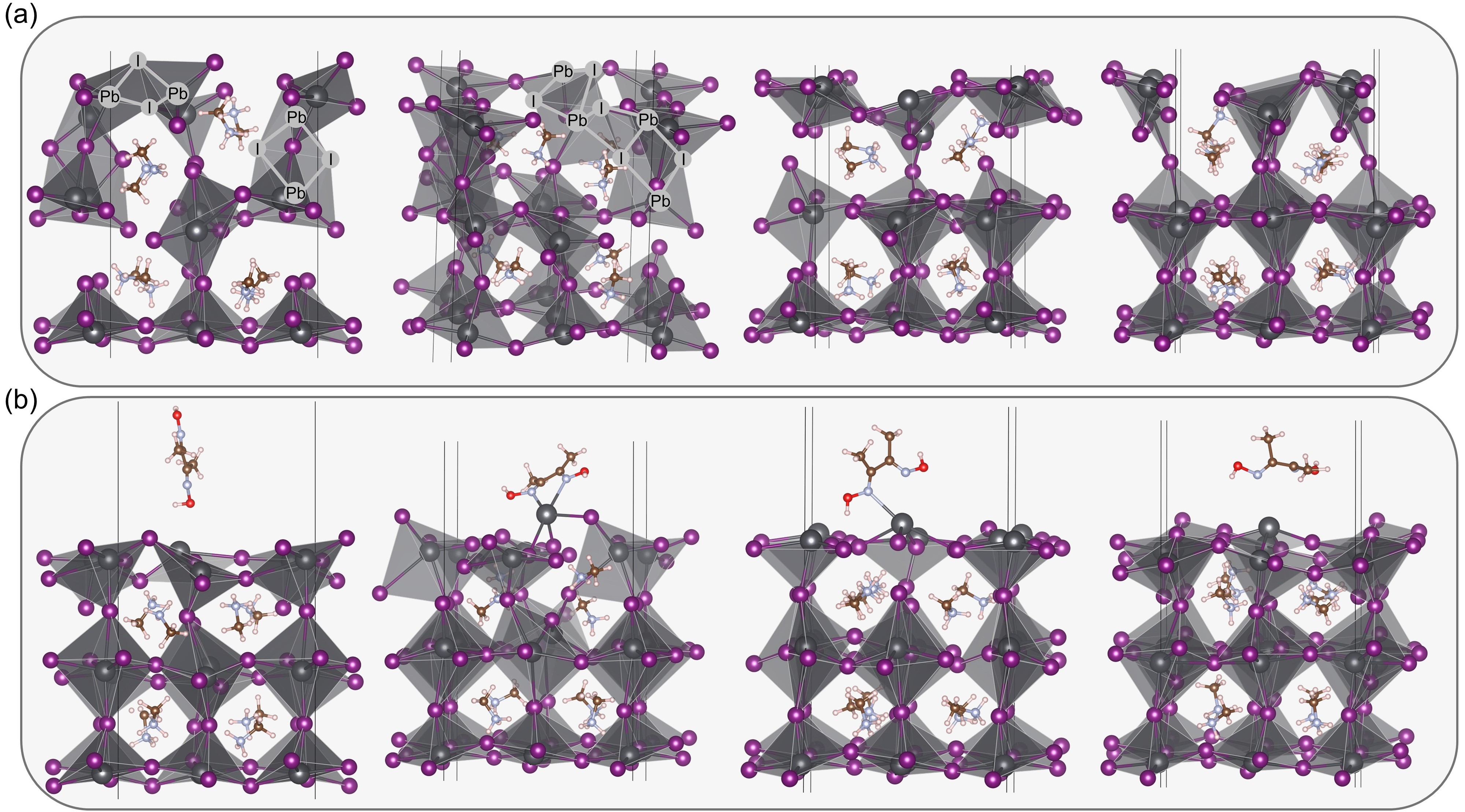}
\renewcommand{\thefigure}{S\arabic{figure}}
\caption{Snapshots at 1.5~ps of four different trajectories under two-hole conditions for (a) pristine MAPbI$_3$ and (b) BDO-passivated MAPbI$_3$.}
\label{fig:1500}
\end{figure}

\begin{figure}[H]
\centering
\includegraphics[width=1\textwidth]{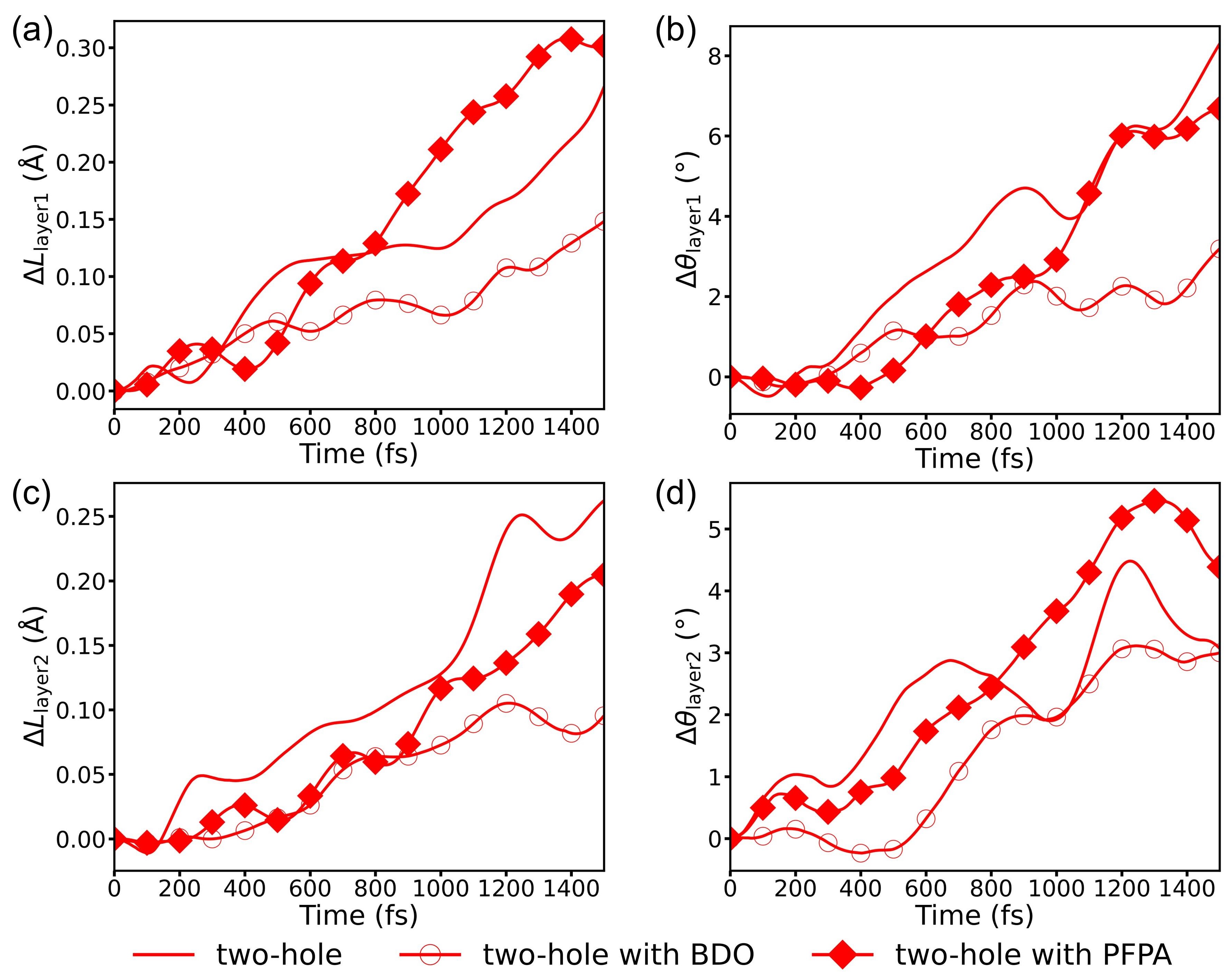}
\renewcommand{\thefigure}{S\arabic{figure}}
\caption{Time evolution of $\Delta L$ and $\Delta \theta$ for layer~1 and layer~2. The corresponding changes in MAPbI$_3$ with BDO are represented by lines with circles, and those in MAPbI$_3$ with PFPA are shown by diamonds.}
\label{fig:BDO-PFPA-2hole-vector}
\end{figure}

\begin{figure}[H]
\centering
\includegraphics[width=1\textwidth]{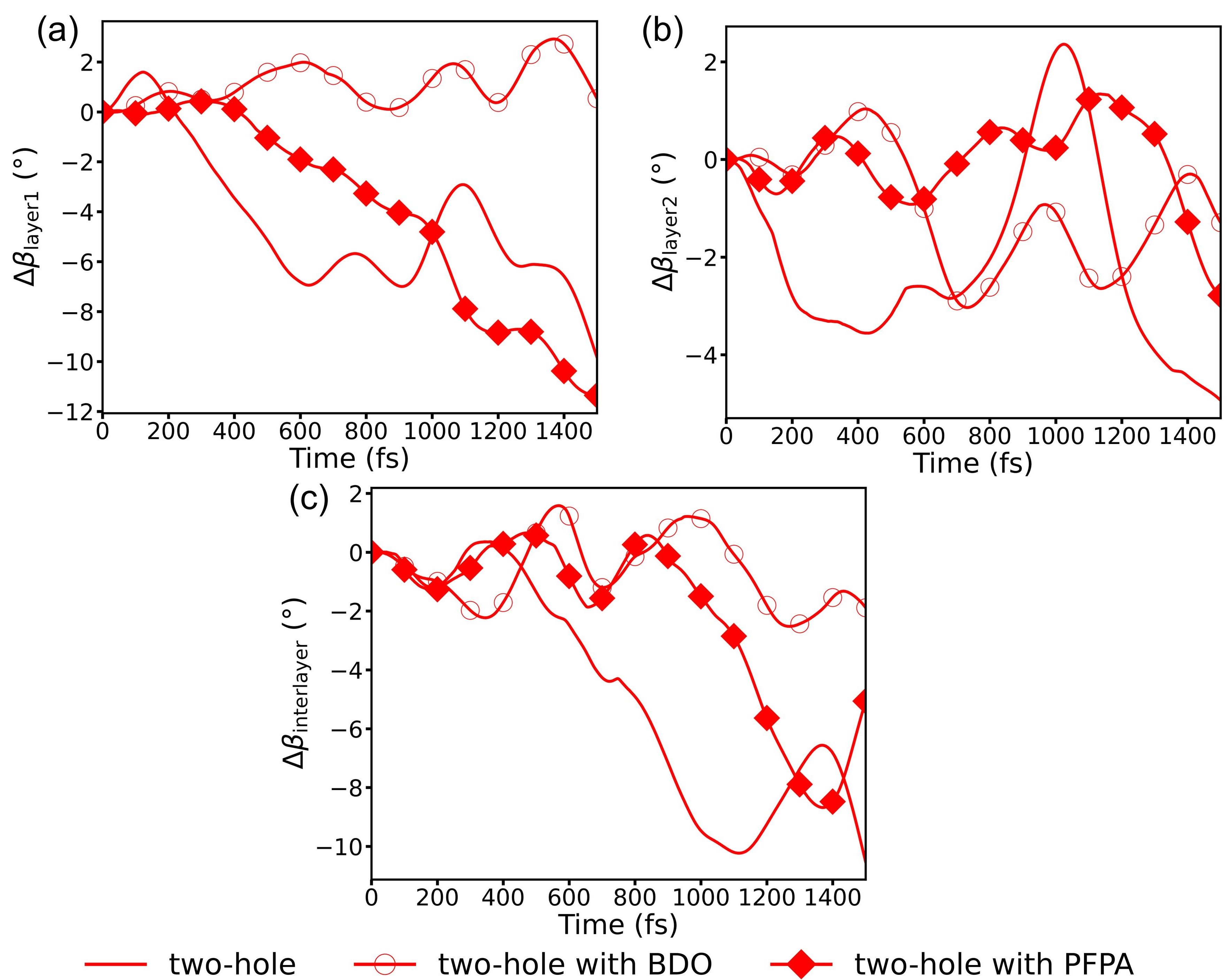}
\renewcommand{\thefigure}{S\arabic{figure}}
\caption{Time evolution of $\Delta \beta$ for layer~1, layer~2, and the interlayer (Pb--I--Pb angle connecting five- and six-coordinated polyhedra). The corresponding changes in MAPbI$_3$ with BDO are represented by lines with circles, and those in MAPbI$_3$ with PFPA are shown by diamonds.}
\label{fig:BDO-PFPA-2hole-angle}
\end{figure}

\begin{figure}[H]
\centering
\includegraphics[width=0.8\textwidth]{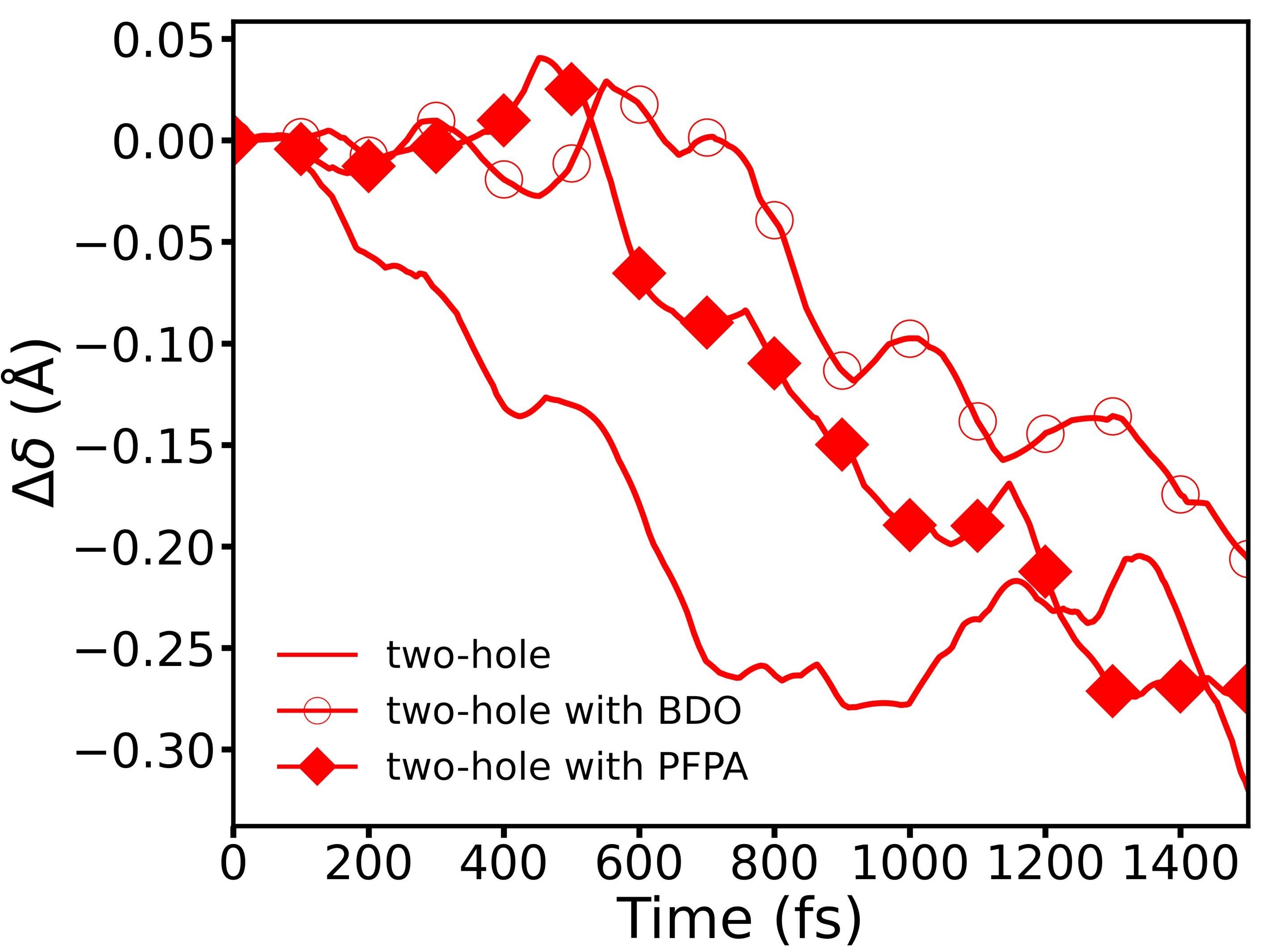}
\renewcommand{\thefigure}{S\arabic{figure}}
\caption{The evolution of $\Delta \delta$. The corresponding changes in MAPbI$_3$ with BDO are represented by lines with circles, and those in MAPbI$_3$ with PFPA are shown by diamonds.}
\label{fig:BDO-PFPA-2hole-min}
\end{figure}

\begin{figure}[H]
\centering
\includegraphics[width=1\textwidth]{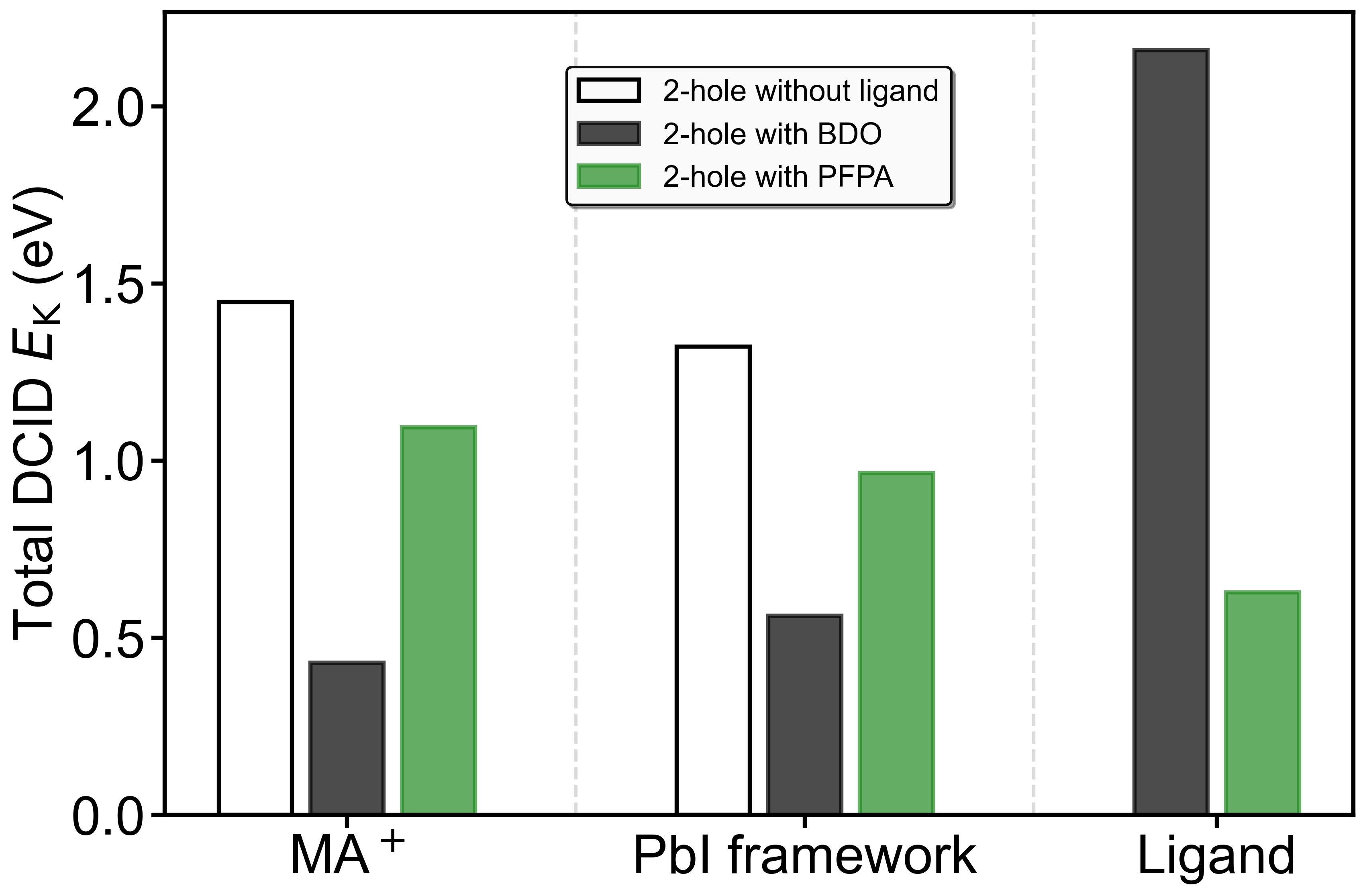}
\renewcommand{\thefigure}{S\arabic{figure}}
\captionof{figure}{The total kinetic energy induced by the NOB event (\ensuremath{E_{k,\text{tot}}}) for different components: the organic part (MA$^+$), the inorganic part (Pb and I), and the passivator. Bars with black and green fill represent the BDO and PFPA systems, respectively.}
\label{fig:Ek_BDO-PFPA}
\end{figure}

\begin{figure}[H]
\centering
\includegraphics[width=0.8\textwidth]{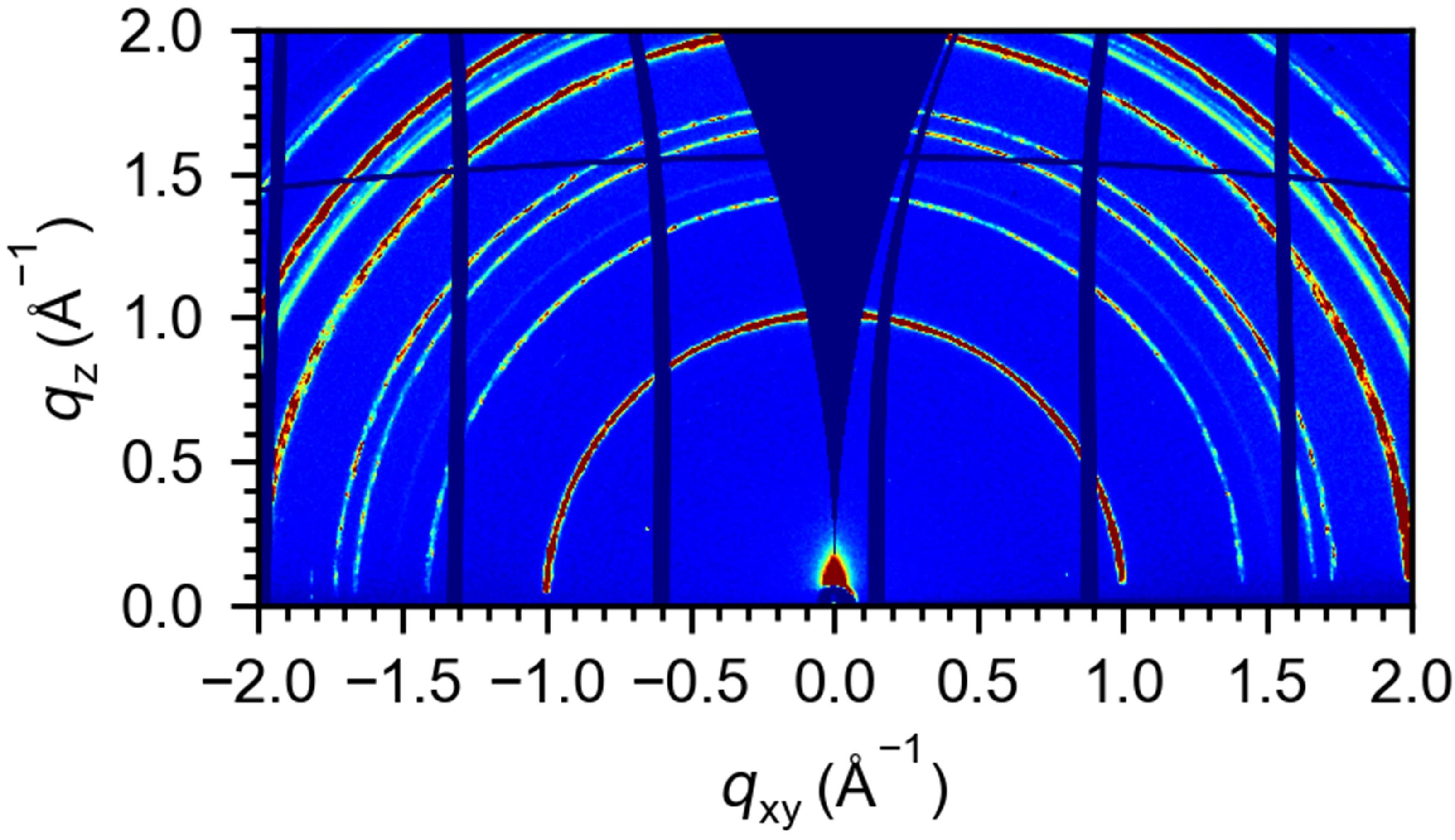}
\renewcommand{\thefigure}{S\arabic{figure}}
\caption{GIWAXS pattern of the MAPbI$_3$ film after 24~h at 50\,$^\circ$C.}
\label{fig:GIWAXS_SI}
\end{figure}

\bibliography{ref}